\documentclass[prb,twocolumn,english,superscriptaddress,amsmath,amssymb,floatfix]{revtex4-2}

\usepackage{float}
\usepackage{graphicx}
\usepackage{dcolumn}
\usepackage{bm}

\usepackage{color}
\usepackage[colorlinks,bookmarks=false,citecolor=darkblue,linkcolor=darkblue,urlcolor=blue]{hyperref}

\definecolor{darkred}{rgb}{0.7,0.0,0.0}

\definecolor{darkblue}{rgb}{0,0.02,0.45}

\definecolor{darkgreen}{rgb}{0.02,0.45,0.0}

\definecolor{violet}{rgb}{0.8,0.2,0.6}

\graphicspath{{graphs/}}

\begin{document}

\title{Boosted magnetic fluctuations at the onset of superconductivity in UTe$_2$ beyond 40 T}

\author{T. Thebault}
 \affiliation{Laboratoire National des Champs Magn\'{e}tiques Intenses - EMFL, CNRS , Univ. Toulouse, Univ. Grenoble Alpes, INSA-T, Toulouse, France}
\author{K. Somesh}
\affiliation{Laboratoire National des Champs Magn\'{e}tiques Intenses - EMFL, CNRS , Univ. Toulouse, Univ. Grenoble Alpes, INSA-T, Toulouse, France}
\author{G. Lapertot}
 \affiliation{Univ. Grenoble Alpes, CEA, Grenoble INP, IRIG, PHELIQS, 38000, Grenoble, France}
 \author{M. Nardone}
\affiliation{Laboratoire National des Champs Magn\'{e}tiques Intenses - EMFL, CNRS , Univ. Toulouse, Univ. Grenoble Alpes, INSA-T, Toulouse, France}
 \author{A. Zitouni}
\affiliation{Laboratoire National des Champs Magn\'{e}tiques Intenses - EMFL, CNRS , Univ. Toulouse, Univ. Grenoble Alpes, INSA-T, Toulouse, France}
  \author{M. Barragan}
\affiliation{Laboratoire National des Champs Magn\'{e}tiques Intenses - EMFL, CNRS , Univ. Toulouse, Univ. Grenoble Alpes, INSA-T, Toulouse, France}
 \author{J. B\'{e}ard}
\affiliation{Laboratoire National des Champs Magn\'{e}tiques Intenses - EMFL, CNRS , Univ. Toulouse, Univ. Grenoble Alpes, INSA-T, Toulouse, France}
 \author{J. Billette}
\affiliation{Laboratoire National des Champs Magn\'{e}tiques Intenses - EMFL, CNRS , Univ. Toulouse, Univ. Grenoble Alpes, INSA-T, Toulouse, France}
 \author{F. Lecouturier-Dupouy}
\affiliation{Laboratoire National des Champs Magn\'{e}tiques Intenses - EMFL, CNRS , Univ. Toulouse, Univ. Grenoble Alpes, INSA-T, Toulouse, France}
 \author{S. Tardieu}
\affiliation{Laboratoire National des Champs Magn\'{e}tiques Intenses - EMFL, CNRS , Univ. Toulouse, Univ. Grenoble Alpes, INSA-T, Toulouse, France}
\author{D. Aoki}
 \affiliation{Institute for Materials Research, Tohoku University, Ikaraki 311-1313, Japan}
\author{G. Knebel}
 \affiliation{Univ. Grenoble Alpes, CEA, Grenoble INP, IRIG, PHELIQS, 38000, Grenoble, France}
\author{D. Braithwaite}
 \affiliation{Univ. Grenoble Alpes, CEA, Grenoble INP, IRIG, PHELIQS, 38000, Grenoble, France}
\author{W. Knafo}
\affiliation{Laboratoire National des Champs Magn\'{e}tiques Intenses - EMFL, CNRS , Univ. Toulouse, Univ. Grenoble Alpes, INSA-T, Toulouse, France}

\date{\today}

\begin{abstract}

Several unconventional superconducting phases have been discovered close to a metamagnetic transition in the heavy-fermion compound UTe$_2$. Although suspected to be of magnetic nature, the mechanisms stabilizing these superconducting phases remain mysterious. Here, we present electrical-resistivity measurements on UTe$_2$, with a current $\mathbf{I}\parallel\mathbf{a}$ and under pulsed magnetic fields up to 60~T rotating in the ($\mathbf{b}$,$\mathbf{c}$) plane. We find that the maximum of the Fermi-liquid coefficient $A$ at the metamagnetic transition is enhanced under magnetic fields tilted by $30-40~^\circ$ from $\mathbf{b}$ to $\mathbf{c}$. The enhancement of $A$ coincides with the stabilization of superconductivity in the polarized paramagnetic regime beyond the metamagnetic field $\mu_0H_m\gtrsim40$~T. It is the signature of a boosted quantum-critical magnetic-fluctuation mode probably in play for the mechanism of this superconducting phase. This result appeals for descriptions of the interplay between magnetic-field-induced superconductivity and quantum critical magnetic properties.

\end{abstract}

\maketitle

In heavy-fermion systems, a Fermi-liquid behavior associated with a large effective mass of quasi-particles is observed and a relation between the amplitude of the effective mass and the strength of magnetic fluctuations has been emphasized \cite{Knafo2021c}. One can indirectly characterize the evolution of the magnetic-fluctuations strength by studying the Fermi-liquid coefficients extracted from thermodynamic or transport experiments, like the Sommerfeld coefficient $\gamma$ from heat-capacity measurements, or the quadratic-temperature-dependence coefficient $A$ from electrical-resistivity measurements. By tuning a parameter, such as chemical doping, pressure or magnetic field, magnetic fluctuations can become quantum critical in the vicinity of a magnetic quantum phase transition, where their enhancement is a precursor of long-range magnetic ordering (see for instance \cite{Knafo2009}). Enhanced quantum critical magnetic fluctuations can also be a driving force for the formation of an unconventional superconducting phase appearing in the vicinity of a magnetic quantum phase transition \cite{Pfleiderer2009}. In the last two decades, magnetic-field-induced superconductivity in the uranium-based heavy-fermion ferromagnets URhGe, UCoGe, and UGe$_2$ has attracted considerable attention \cite{Aoki2019b}. In these systems, a superconducting phase is induced or reinforced in a magnetic field and develops in the vicinity of a metamagnetic transition, at which ferromagnetic fluctuations are suspected to be involved in a triplet superconducting mechanism \cite{Hattori2012,Wu2017}. In URhGe, an enhancement of the Fermi-liquid coefficient $A$ at the metamagnetic transition, in the vicinity of a field-induced superconducting phase, was proposed to result from ferromagnetic fluctuations \cite{Miyake2008,Miyake2009,Gourgout2016}. The presence of such quantum critical magnetic fluctuations at the metamagnetic transition was also supported by nuclear-magnetic-resonance (NMR) relaxation-rate studies \cite{Tokunaga2015,Tokunaga2016}. More recently, several field-induced superconducting phases have been discovered in the heavy-fermion paramagnet UTe$_2$, revealing a new playground to study the relation between magnetism and superconductivity, with possible triplet mechanism \cite{Ran2019a,Aoki2019a,Ran2019b,Knebel2019,Aoki2021,Valiska2021,Kinjo2023,Frank2024,Wu2025,Lewin2025}.

\begin{figure*}[!hbt]
\centering
\includegraphics[width=0.8\textwidth]{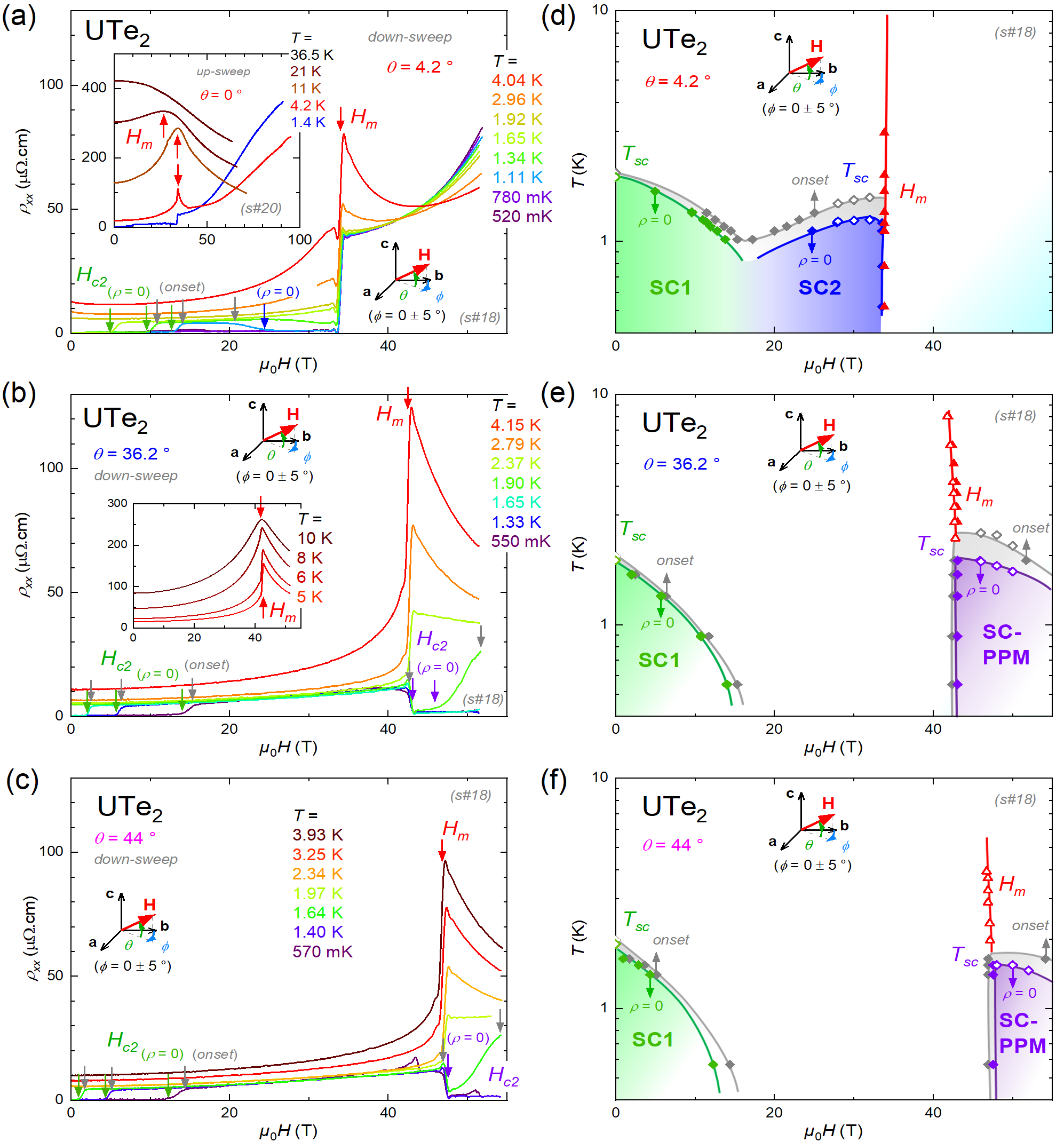}
\caption{\label{Fig1} Left-hand graphs: Electrical resistivity $\rho_{xx}$ of UTe$_2$ versus magnetic field up to 55~T, at temperatures $T$ from 500~mK to 36.5~K, for field directions with the angles $\theta=(\mathbf{b},\mathbf{H})=$ (a) $4.2~^\circ$, (b) $36.2~^\circ$, and (c) $44~^\circ$. The data in the main Panels and in the Inset of Panel (b) were obtained on sample $\#18$ during the fall of pulsed magnetic fields up to 55~T, and the data in the Inset of Panel (a) were obtained on sample $\#20$ during the rise of pulsed magnetic fields up to 95~T. Right-hand graphs: Magnetic-field-temperature phase diagrams obtained for $\theta=$ (d) $4.2~^\circ$, (e) $36.2~^\circ$, and (f) $44~^\circ$. In the phase diagrams, the open symbols correspond to points extracted from $\rho_{xx}(T)$ and closed symbols correspond to points extracted from $\rho_{xx}(H)$.}
\end{figure*}

UTe$_2$ is a nearly antiferromagnet characterized by strong antiferromagnetic fluctuations \cite{Duan2020,Knafo2021b,Butch2022,Duan2021,Raymond2021,Tokunaga2019,Fujibayashi2023,Matsumura2025}, and it can be tuned to long-range antiferromagnetic order via a quantum magnetic phase transition under pressure \cite{Knafo2025}. At zero magnetic field, the superconducting phase SC1 is stabilized below $T_{sc} \approx$~2~K at ambient pressure, and a second superconducting phase SC2 develops below $T_{sc} \approx$~3~K near the quantum magnetic phase transition under pressure \cite{Braithwaite2019}. At ambient pressure, for $T\rightarrow0$, and in a magnetic field $\mathbf{H}$ along the direction $\mathbf{b}$, SC1 vanishes near 20~T and is replaced by SC2, which is induced near a metamagnetic transition at $\mu_0H_m=34$~T, beyond which SC2 vanishes \cite{Knebel2019,Ran2019b,Knafo2019,Miyake2019,Rosuel2023a,Vasina2025}. By tilting the magnetic field from $\mathbf{b}$ to $\mathbf{c}$, a second field-induced superconducting phase SC-PPM is stabilized in the polarized paramagnetic (PPM) regime for $H\gtrsim H_m$, with $\mu_0H_m=40-45$~T \cite{Ran2019b,Knafo2021a,Helm2024}. The metamagnetic field $H_m$ and the boundaries of SC2 and SC-PPM were mapped out as function of the three components of the magnetic-field direction, showing a complex three-dimensional phase diagram \cite{Wu2025,Lewin2025,Lewin2024}. The Fermi-liquid Sommerfeld coefficient $\gamma$ in the heat capacity and the quadratic coefficient $A$ in the electrical resistivity of UTe$_2$ were shown to exhibit a maximum at the metamagnetic transition for the two field directions $\mathbf{H}\parallel\mathbf{b}$ and $\mathbf{H}$ tilted by $\approx 30-40~^\circ$ from $\mathbf{b}$ to $\mathbf{c}$ \cite{Rosuel2023a,Knafo2019,Knafo2021a,Imajo2019,Miyake2021,Thebault2022}. For $\mathbf{H}\parallel\mathbf{b}$, a similar magnetic-field variation of the NMR relaxation rates and electrical-resistivity coefficient $A$ (measured with an electrical current $\mathbf{I}\parallel\mathbf{a}$) was found, supporting that the Fermi-liquid regime is controlled by quantum critical magnetic fluctuations peaked at the metamagnetic transition \cite{Tokunaga2023}. However, the reasons for the different domains of stability of SC2 and SC-PPM remain mysterious. A systematic characterization of the quantum critical magnetic fluctuations developing near the metamagnetic transition, under high magnetic fields rotating in the ($\mathbf{b}$,$\mathbf{c}$) plane, is now needed to address this question.

In this letter, we present electrical-resistivity measurements on UTe$_2$ single crystals in magnetic fields applied in the ($\mathbf{b}$,$\mathbf{c}$) plane. Different samples (labeled $\#16$, $\#18$, $\#19$, and $\#20$; see details in the Supplemental Material \cite{SM} - see also references \cite{Wu2024,Aoki2024,Knafo2017,Sakai2023,Tokiwa2023} therein) grown by the molten-salt-flux (MSF) method \cite{Sakai2022} have been studied. The electrical resistivity $\rho_{xx}$ has been measured by the four-point method with a current $\mathbf{I}\parallel\mathbf{a}$. Magnetic fields have been generated by long-duration pulsed magnets at the Laboratoire National des Champs Magn\'{e}tiques Intenses in Toulouse. A standard single-coil magnet has been used for most of the measurements presented here, which were done in magnetic fields up to 55~T and temperatures from 500~mK to 10~K delivered by a 10-mm bore $^3$He insert. A triple-coil prototype was used to generate magnetic fields up to 95~T (see \cite{Beard2018}), at temperatures from 1.4 to 36.5~K delivered by a 4-mm bore $^4$He cryostat, for complementary measurements. We have extracted the magnetic-field evolution of the Fermi-liquid coefficient $A$ and of the superconducting temperatures $T_{SC2}$ and $T_{SC-PPM}$ of the phases SC2 and SC-PPM, respectively, for field angles $\theta=(\mathbf{b},\mathbf{H})$ varying from from 0 to $46~^\circ$.  Our data show signatures of a boosted magnetic-fluctuations mode which may play a role in the mechanism driving SC-PPM.

The electrical resistivity $\rho_{xx}$ measured in magnetic fields with the angles $\theta=4.2$, 36.2 and 44$~^\circ$ and temperatures from $T=500$~mK to 10~K is presented in Figs. \ref{Fig1}(a-c). At $\theta=4.2~^\circ$, SC2 is stabilized in the field window $15$~T~$<\mu_0H<\mu_0H_m\simeq34$~T, at temperatures $T \le$~1.2~K [Fig. \ref{Fig1}(a)]. At temperatures $T\le$~6~K, a sharp step-like increase of the resistivity occurs at the metamagnetic field $H_m$. At temperatures $T>6$~K, the metamagnetic transition is transformed into a crossover characterized by a broad maximum in $\rho_{xx}$, as observed in previous studies \cite{Knafo2019,Knafo2021a}. An increase of the resistivity $\rho_{xx}$ is observed at high fields and low temperatures, from $\simeq45~\mu\Omega$cm at $\mu_0H=45$~T to $\simeq350~\mu\Omega$cm at $\mu_0H=95$~T at the temperature $T=1.4$~K [see Inset of Fig. \ref{Fig1}(a)]. This increase of $\rho_{xx}$ is attributed to the cyclotron motion of the conduction electrons in a transverse configuration (current orthogonal to the magnetic field) \cite{Aoki2022a,Onuki2018}. A similar behavior was already observed in the electrical resistivity of UTe$_2$ with electrical currents $\mathbf{I}\parallel\mathbf{a},\mathbf{c}$ under a magnetic field $\mathbf{H}\parallel\mathbf{b}$ \cite{Knafo2019,Thebault2022}. For $\theta=36.2$ and $44~^\circ$, the superconducting phase SC-PPM is stabilized for $H>H_m$ with $\mu_0H_m=43$~T and 47.5~T, respectively [Figs. \ref{Fig1}(b-c)]. No cyclotron-motion effect is observed in fields up to 60~T applied along these directions. In the vicinity of $H_m$, the temperature variation of the electrical resistivity is the largest for $\theta=36.2~^\circ$. Magnetic-field-temperature phase diagrams obtained from our electrical-resistivity measurements are presented in Figs. \ref{Fig1}(d-f). For $\theta=4.2~^\circ$, the highest critical temperature $T_{SC2}^{max}=1.25$~K of SC2 is found at $\mu_0H \approx$~32~T, i.e., just below $H_m$ [Fig. \ref{Fig1}(d)]. For $\theta=36.2$ and 44$~^\circ$, the critical temperature of SC-PPM is maximum in fields just above the metamagnetic field, where it reaches $T_{SC-PPM}^{max}=1.9$~K and 1.55~K, respectively [Figs. \ref{Fig1}(e-f)].

\begin{figure}[!t]
\centering
\includegraphics[width=0.45\textwidth]{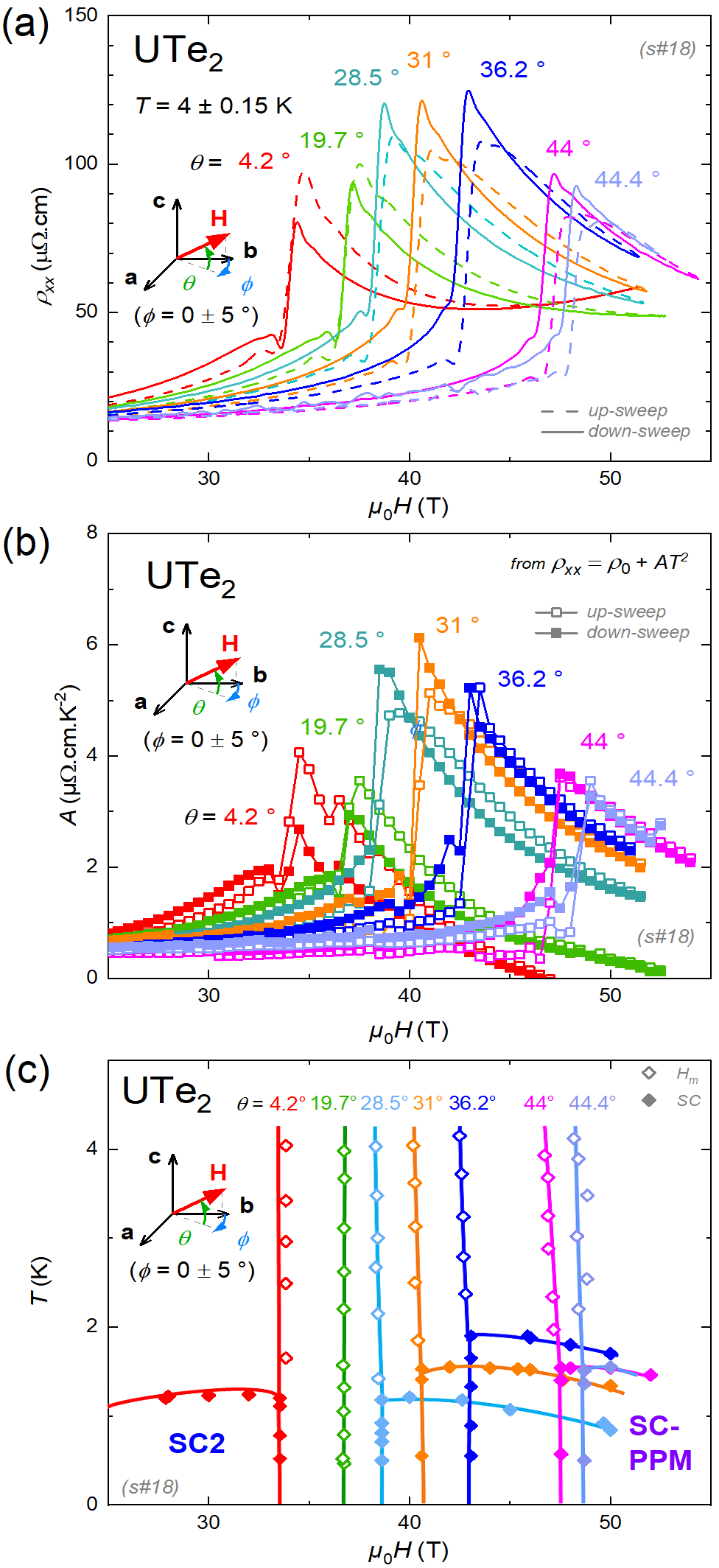}
\caption{\label{Fig2} Magnetic-field variations of (a) the electrical resistivity $\rho_{xx}$ at $T=4\pm0.15$~K, (b) of the Fermi-liquid coefficient $A$, and (c) magnetic-field-temperature phase diagrams of UTe$_2$ under a magnetic field oriented with angles $\theta$ from 4.2 to $44.4~^\circ$. Data from up- and down-sweeps of the pulsed field are shown in panels (a,b).}
\end{figure}

The electrical resistivity $\rho_{xx}$ measured at $T=4\pm0.15$~K in magnetic fields with angles $\theta$ varying from $4.2~^\circ$ to $44.4~^\circ$ is presented in Fig. \ref{Fig2}(a). Small hysteresis loops (due to the first-order nature of the transition at $H_m$, and possibly additional magnetocaloric effects \cite{Schonemann2023} and eddy-current heating) visible in our up- and down-sweep field data do not affect the observations and analyzes made in the following. $\rho_{xx}$ is maximum at the metamagnetic transition and its maximum is enhanced at angles $28.5\leq\theta\leq36.2~^\circ$, where it reaches $\simeq120~\mu\Omega$cm and where the largest resistivity jump $\Delta\rho(H_m)\simeq70~\mu\Omega$cm is observed at $H_m$. Fig. \ref{Fig2}(b) shows the magnetic-field dependence of the Fermi-liquid coefficient $A$ extracted from fits by $\rho_{xx}=\rho_0 + AT^2$ to our data (see Supplemental Material \cite{SM}). For each magnetic-field direction, the coefficient $A$ reaches its maximal value $A^{max}$ at $H_m$, and $A^{max}$ is enhanced at $\theta\simeq30-35~^\circ$. At angles $\theta\geq20~^\circ$, the maximum of $A$ at $H_m$ is strongly asymmetric, with a step-like increase for $H \lesssim H_m$ and a shoulder for $H \gtrsim H_m$. The $H$ and $\theta$ variations of $A$ shown here for sample $\#18$ are consistent with those extracted for other samples grown by the MSF and chemical-vapor-transport (CVT) techniques (see \cite{Knafo2019,Knafo2021a} and Supplemental Material \cite{SM}). We note that, in \cite{Weinberger2025}, it was stressed that SC-PPM forms out of a 'strange metal' with a linear temperature dependence of the electrical resistivity. However, for the data presented here, an analysis by $\rho_{xx}=\rho_0 + A_nT^n$ with a free coefficient $n$ is not relevant since it leads to non-physical negative values of $\rho_0$ with $n\simeq1$ fitted near $H_m$ \cite{Knafo2026}.  Fig. \ref{Fig2}(c) shows a superposition of the magnetic-field-temperature phase diagrams obtained for the different field directions. For $\theta=4.2~^\circ$, SC2 is fully stabilized in fields $H<H_m$. For $\theta=19.7~^\circ$, the onset of SC2 is observed in fields $H<H_m$ but zero resistivity is not reached at low temperature. For $\theta >19.7~^\circ$, SC-PPM is stabilized in fields $H>H_m$. The critical temperature of SC-PPM reaches its maximum value $T_{SC-PPM}^{max}\simeq1.9$~K for $\theta=36.2~^\circ$.

\begin{figure}[t]
\centering
\includegraphics[width=0.5\textwidth]{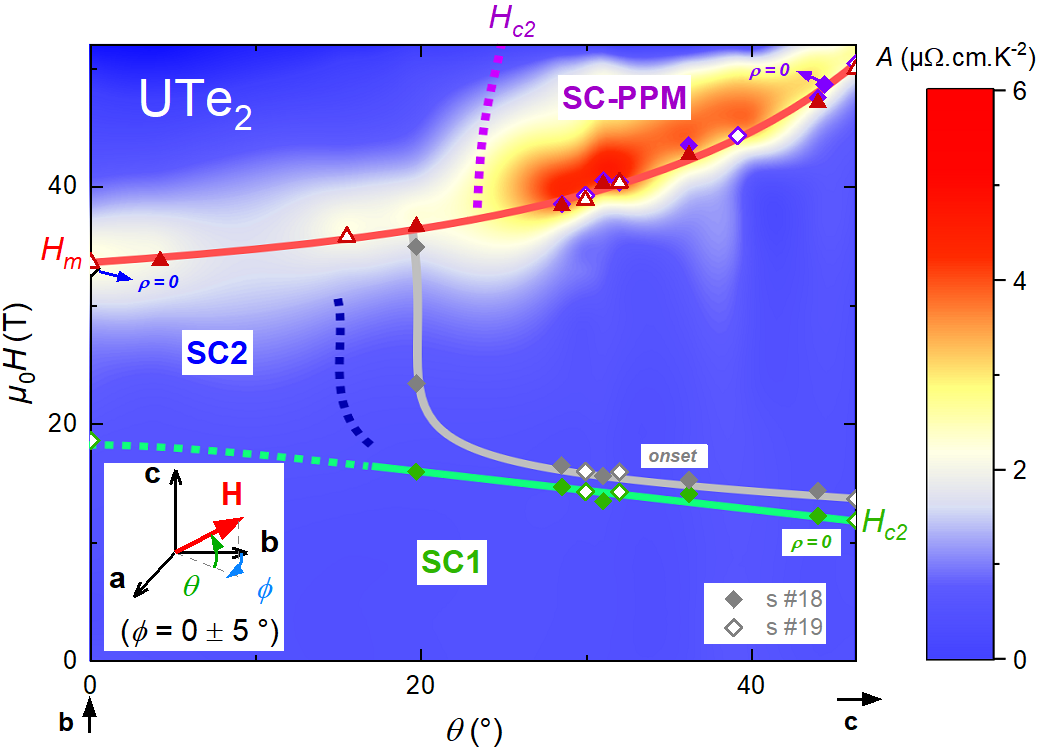}
\caption{\label{Fig3} Intensity map of the Fermi-liquid coefficient $A$ as a function of $H$ and $\theta=(\mathbf{b},\mathbf{H})$ (for $\mathbf{H}\perp\mathbf{a}$). The intensity map combines down-sweep data from samples $\#18$ and $\#19$ (see details in the Supplemental Material \cite{SM}). The phase diagram shown on top of the intensity map was obtained on samples $\#18$ and $\#19$ at $T=500$~mK. The dashed lines indicate expected transition lines which were not probed here.}
\end{figure}

Figure \ref{Fig3} shows an intensity map of $A$ as a function of $\theta$ and $H$, combining down-sweep data from samples $\#18$ (presented here) and $\#19$ (see Supplemental Material \cite{SM}). The phase diagram of UTe$_2$ shown on top of the contour plot was obtained from data collected on samples $\#18$ and $\#19$ at $T \simeq500$~mK, the $H_{c2}$ line of SC-PPM being taken from \cite{Helm2024,Ran2019b}. The angle dependencies of the maximum in field $A^{max}$ of the Fermi-liquid coefficient (collected on samples $\#16$, $\#18$ and $\#19$ here, and on samples  $\#1$ to $\#7$ in \cite{Knafo2019,Knafo2021a}) and of the critical temperatures $T_{SC2}^{max}$ and $T_{SC-PPM}^{max}$ of SC2 and SC-PPM (collected on samples $\#16$, $\#18$ and $\#19$ here, and from \cite{Knebel2019,Helm2024}) are presented in Figs. \ref{Fig4}(a,b), respectively (see Supplemental Material \cite{SM}). These plots show that $A^{max}$ is enhanced at angles $\theta\simeq30-35~^\circ$ while $T_{SC-PPM}^{max}$ is enhanced at angles $\theta\simeq35-39~^\circ$.

\begin{figure}[t]
\centering
\includegraphics[width=0.45\textwidth]{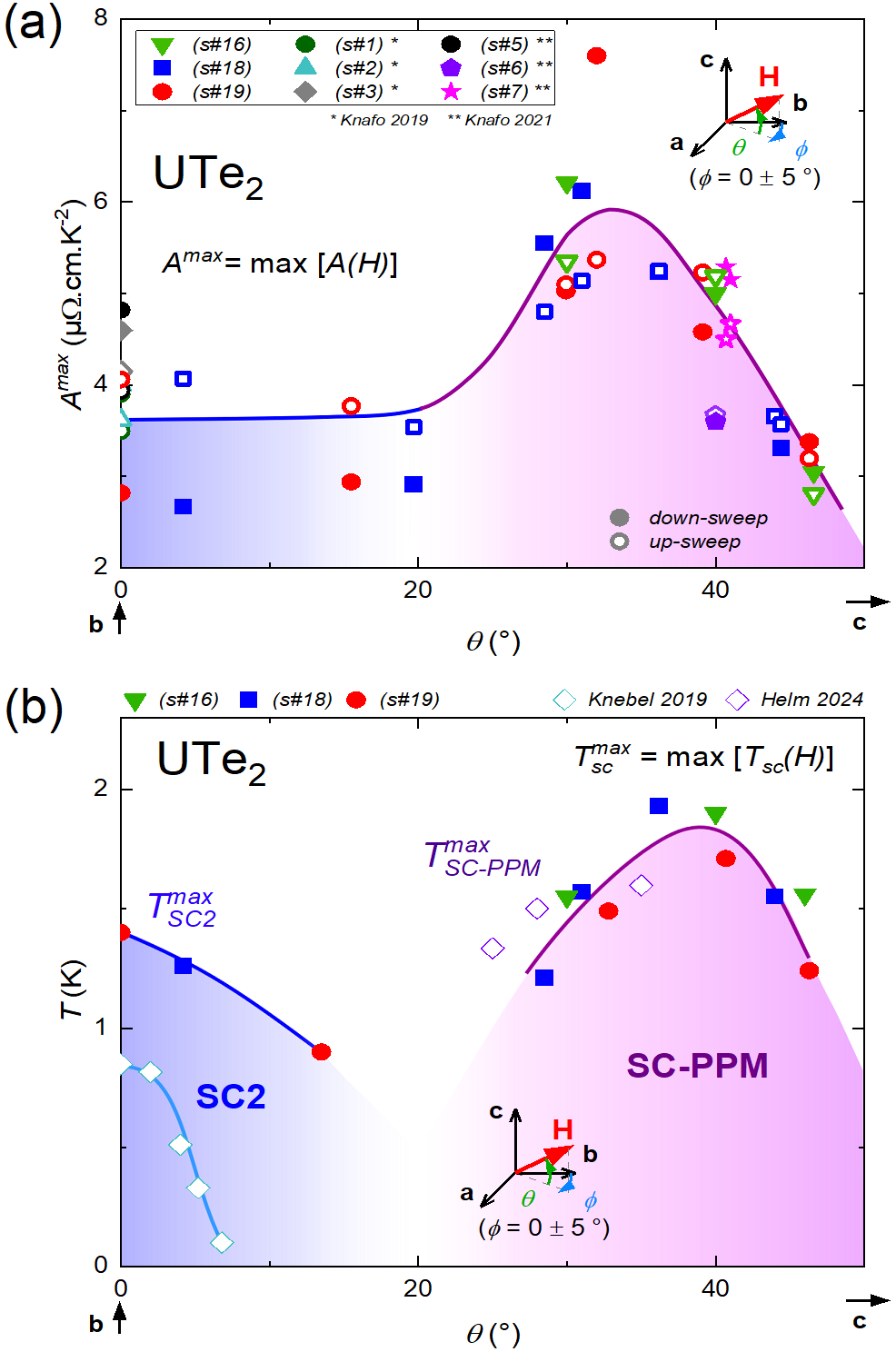}
\caption{\label{Fig4} (a) Angle-$\theta$ variation of $A^{max}=max[A(H)]$ extracted from the electrical resistivity of the MSF samples $\#16$, $\#18$ and $\#19$ investigated here and on CVT samples $\#1$, $\#2$, $\#3$, $\#5$, $\#6$ and $\#7$ studied in \cite{Knafo2019,Knafo2021a} (data from up- and down-sweeps of the pulsed field are shown). (b) Angle-$\theta$ variation the maximal critical temperatures $T_{SC2}^{max}$ and $T_{SC-PPM}^{max}$ of the superconducting phases SC2 and SC-PPM, respectively, extracted on the MSF samples $\#16$, $\#18$ and $\#19$ studied here and on CVT samples studied in \cite{Knebel2019,Helm2024}. The blue and violet background colors indicate the regimes where the phases SC2 and SC-PPM, respectively, are stabilized in high field.}
\end{figure}

While quantum critical magnetic fluctuations are expected near a second-order quantum magnetic phase transition \cite{Hertz1976,Millis1993,Moriya1995}, the presence of magnetic fluctuations in the vicinity of a first-order quantum magnetic phase transition was scarcely considered. In the prototypical heavy-fermion antiferromagnetic system Ce$_{1-x}$La$_x$Ru$_2$Si$_2$, critical antiferromagnetic fluctuations were identified near a quantum magnetic phase transition suspected to be first-order \cite{Knafo2009}. In UTe$_2$ under magnetic field $\mathbf{H}\parallel\mathbf{b}$, the metamagnetic transition also turns first-order at low temperatures \cite{Miyake2019,Miyake2021} and the presence of enhanced magnetic fluctuations near $H_m$, first indicated by electrical-resistivity measurements \cite{Knafo2019,Knafo2021a,Thebault2022}, was confirmed by NMR relaxation-rate measurements \cite{Tokunaga2023}. Here, we find similar variations of the Fermi-liquid coefficient $A$ [Fig. \ref{Fig2}(a)] and $\rho_{xx}$ at $T=4$~K [Fig. \ref{Fig2}(b)] resulting from the contribution of magnetic fluctuations. The entire $H_m$ versus $\theta$ line is accompanied by quantum critical magnetic fluctuations, as indicated by the maximum value $A^{max}$ of $A$ at $H_m$ observed for all values of $\theta$. We further find that enhanced values of $A^{max}$ coincide with the highest superconducting temperature $T_{SC-PPM}^{max}$ of SC-PPM found near $H_m$ at similar angles $30\lesssim\theta\lesssim40~^\circ$. This is the signature of a boosted quantum critical mode of magnetic fluctuations probably in play in the mechanism driving to SC-PPM. On the contrary, at low angles $\theta<20~^\circ$, $A^{max}$ is almost constant while the highest superconducting temperature $T_{SC2}^{max}$ of SC2 rapidly decreases with $\theta$, and their variations in angle seem not to be simply correlated. The increase of $A^{max}$ seems then too slow to overcome the superconducting-pair breaking induced by the magnetic field. To allow superconductivity at higher fields, larger values of $A^{max}$ than those found for $\mathbf{H}\simeq\parallel\mathbf{b}$ would even be needed to overcome the breaking effect, and larger values of $A^{max}$ are indeed found here at angles $30\lesssim\theta\lesssim40~^\circ$. However, the situation is not so simple, since similar values of $T_{SC-PPM}^{max}$ are found at angles $\theta=31$ and 44.4~$^\circ$, but $A^{max}$ is smaller for $\theta=44.4~^\circ$ than for $\theta=31~^\circ$.

Interestingly, large values of $A$ extracted from $\rho_{xx}$ are found, with an asymmetric shape, above $H_m$, at angles $30\lesssim\theta\lesssim40~^\circ$ (see Fig. \ref{Fig2}(a) and Fig. S14 in the Supplemental Material \cite{SM}). The $\theta$-variation of $A$ for $H>H_m$ reflects that of its field-maximum value $A_{max}$. This may qualitatively explain the angular range for SC-PPM and the fact that it only occurs above $H_m$. For $H<H_m$, smaller values of $A$ are found and, at constant magnetic field, the decrease of $A$ with increasing angle $\theta$ may be correlated with the suppression of SC2, which is observed for $H<H_m$ and $\theta\approx0~^\circ$. This indicates that $A$ may also be, in a lesser degree, sensitive to a magnetic-fluctuation mode implied in the mechanism of SC2. Such asymmetry in the variation of $A$ through $H_m$ was already observed for an angle $\theta=41\pm2~^\circ$ (initially estimated at $27\pm5~^\circ$, see Supplemental Material \cite{SM}) in \cite{Knafo2021a}. A similar asymmetry was also found in the Sommerfeld coefficient extracted, using thermodynamic relationships, from magnetization measurements performed with an angle $\theta\simeq28~^\circ$ \cite{Miyake2021}. On the contrary, a more symmetric variation of $A$ extracted from $\rho_{xx}$ through $H_m$  is observed in configurations where SC2 is stabilized for $H<H_m$, either for $\mathbf{H}\parallel\mathbf{b}$ and at small angles $\theta$ (see here and in \cite{Knafo2019,Knafo2021a}), or under pressure close to $p_c\simeq1.5-1.7$~GPa combined with a magnetic field tilted by $\theta\simeq30~^\circ$ \cite{Valiska2021,Thebault2024}.

In the prototypical heavy-fermion paramagnet CeRu$_2$Si$_2$, the metamagnetic transition is accompanied by a maximum of the magnetic fluctuations at \cite{Flouquet2004,Ishida1998,vanderMeulen1991,Aoki2011b} and a change of the Fermi surface \cite{Takashita1996,Boukahil2014} (see also \cite{Knafo2021c}). In UTe$_2$, a step-like increase of the residual resistivity $\rho_0$ - extracted from the Fermi-liquid fits - at $H_m$ (see the Supplemental Material \cite{SM}) may result from the polarization of the magnetic moments combined with a change of the Fermi surface. In addition to the effects from quantum-critical magnetic fluctuations, we may expect that SC2 stabilized for $H<H_m$ and SC-PPM stabilized for $H>H_m$ are associated with different Fermi surfaces. A Fermi-surface change at $H_m$ was suggested from Hall-effect experiments in magnetic fields $\mathbf{H}\parallel\mathbf{b}$ \cite{Niu2020b}. A characterization by quantum-oscillation measurements of the Fermi surface in fields beyond $H_m$, in addition to those already done for $H<H_m$ \cite{Aoki2022a,Eaton2024}, would be needed to describe SC2 and SC-PPM. We note that, while the Fermi surface is presumably reconstructed at $H_m$, it is not expected to be modified for $H<H_m$ where the coefficient $A$, which is mainly driven by the magnetic fluctuations, slowly increases.

Further experiments on UTe$_2$ may help accessing new information about the quantum critical magnetic-fluctuations modes and their relation with the domains of stability of SC2 and SC-PPM. The magnetic fluctuations in UTe$_2$, as in other U-based heavy-fermion systems, are a property of $5f$ electrons, which  contribute to the Fermi surface and, therefore, to the electrical transport. For actinide metals and compounds, the effect of magnetic fluctuations to the electrical resistivity was modeled assuming isotropic properties in \cite{Jullien1974}. However, the electrical resistivity of UTe$_2$ strongly depends on the electrical-current direction \cite{Eo2022,Thebault2022,Knebel2024} and one can suspect that its anisotropy may, at least partly, be related with the anisotropy of the magnetic fluctuations. A further investigation by electrical resistivity with currents $\mathbf{I}\parallel\mathbf{b},\mathbf{c}$ may permit accessing complementary information to that extracted here with a current $\mathbf{I}\parallel\mathbf{a}$. A heat-capacity study would allow extracting the field- and angle-variations of the Sommerfeld coefficient, which gives a direct information about the entropy associated with the magnetic fluctuations. NMR relaxation-rate studies may also be performed to extract microscopic information about the magnetic-fluctuations modes implied in the stabilization of SC2 and SC-PPM. Such projets in rotating and intense magnetic fields request pushing up the experimental state of the art and constitute challenges to overcome for the next years.

\section*{Acknowledgements}

This work was performed at the Laboratoire National des Champs Magn\'{e}tiques Intenses, a member of the European Magnetic Field Laboratory. We acknowledge financial support from the French National Research Agency collaborative research projects FRESCO No. ANR-20-CE30-0020 and SCATE No. ANR-22-CE30-0040, and from the JSPS KAKENHI Grants Nos. JP22H04933, JP20KK0061, JP24H01641, JP25H01249.


%

\newpage

\onecolumngrid

\renewcommand\thefigure{S\arabic{figure}}
\renewcommand{\theequation}{S\arabic{equation}}
\renewcommand{\thetable}{S\arabic{table}}
\renewcommand{\bibnumfmt}[1]{[S#1]}
\renewcommand{\citenumfont}[1]{S#1}
\setcounter{figure}{0}
\renewcommand{\thesection}{S\arabic{section}}
\renewcommand{\thesubsection}{S\arabic{subsection}}

\vspace{15cm}

\begin{center}
\large {\textbf {Supplemental Material: \\Boosted magnetic fluctuations at the onset of superconductivity in UTe$_2$ beyond 40 T}}
\end{center}
\vspace{1cm}

\setcounter{page}{1}

Experimental details are given and supplemental figures are shown for the different investigated samples. Details about raw data and their analysis to construct magnetic-field-temperature phase diagrams and to extract the Fermi-liquid coefficient $A$ are given. We also show that a flux-flow resistive signal persists in the superconducting phase SC1 for $\mathbf{H}\approx\parallel\mathbf{b}$ in these samples grown by the molten-salt-flux technique.

\newpage

\author{T. Thebault}
 \affiliation{Laboratoire National des Champs Magn\'{e}tiques Intenses - EMFL, CNRS , Univ. Toulouse, Univ. Grenoble Alpes, INSA-T, Toulouse, France}
\author{K. Somesh}
\affiliation{Laboratoire National des Champs Magn\'{e}tiques Intenses - EMFL, CNRS , Univ. Toulouse, Univ. Grenoble Alpes, INSA-T, Toulouse, France}
\author{G. Lapertot}
 \affiliation{Univ. Grenoble Alpes, CEA, Grenoble INP, IRIG, PHELIQS, 38000, Grenoble, France}
 \author{M. Nardone}
\affiliation{Laboratoire National des Champs Magn\'{e}tiques Intenses - EMFL, CNRS , Univ. Toulouse, Univ. Grenoble Alpes, INSA-T, Toulouse, France}
 \author{A. Zitouni}
\affiliation{Laboratoire National des Champs Magn\'{e}tiques Intenses - EMFL, CNRS , Univ. Toulouse, Univ. Grenoble Alpes, INSA-T, Toulouse, France}
  \author{M. Barragan}
\affiliation{Laboratoire National des Champs Magn\'{e}tiques Intenses - EMFL, CNRS , Univ. Toulouse, Univ. Grenoble Alpes, INSA-T, Toulouse, France}
 \author{J. B\'{e}ard}
\affiliation{Laboratoire National des Champs Magn\'{e}tiques Intenses - EMFL, CNRS , Univ. Toulouse, Univ. Grenoble Alpes, INSA-T, Toulouse, France}
 \author{J. Billette}
\affiliation{Laboratoire National des Champs Magn\'{e}tiques Intenses - EMFL, CNRS , Univ. Toulouse, Univ. Grenoble Alpes, INSA-T, Toulouse, France}
 \author{F. Lecouturier-Dupouy}
\affiliation{Laboratoire National des Champs Magn\'{e}tiques Intenses - EMFL, CNRS , Univ. Toulouse, Univ. Grenoble Alpes, INSA-T, Toulouse, France}
 \author{S. Tardieu}
\affiliation{Laboratoire National des Champs Magn\'{e}tiques Intenses - EMFL, CNRS , Univ. Toulouse, Univ. Grenoble Alpes, INSA-T, Toulouse, France}
\author{D. Aoki}
 \affiliation{Institute for Materials Research, Tohoku University, Ikaraki 311-1313, Japan}
\author{G. Knebel}
 \affiliation{Univ. Grenoble Alpes, CEA, Grenoble INP, IRIG, PHELIQS, 38000, Grenoble, France}
\author{D. Braithwaite}
 \affiliation{Univ. Grenoble Alpes, CEA, Grenoble INP, IRIG, PHELIQS, 38000, Grenoble, France}
\author{W. Knafo}
\affiliation{Laboratoire National des Champs Magn\'{e}tiques Intenses - EMFL, CNRS , Univ. Toulouse, Univ. Grenoble Alpes, INSA-T, Toulouse, France}

\date{\today}

\renewcommand\thefigure{S\arabic{figure}}
\renewcommand{\theequation}{S\arabic{equation}}
\renewcommand{\thetable}{S\arabic{table}}
\setcounter{figure}{0}
\onecolumngrid

\subsection{Experimental details}

\begin{figure*}[tb]
\centering
\includegraphics[width=0.8\textwidth]{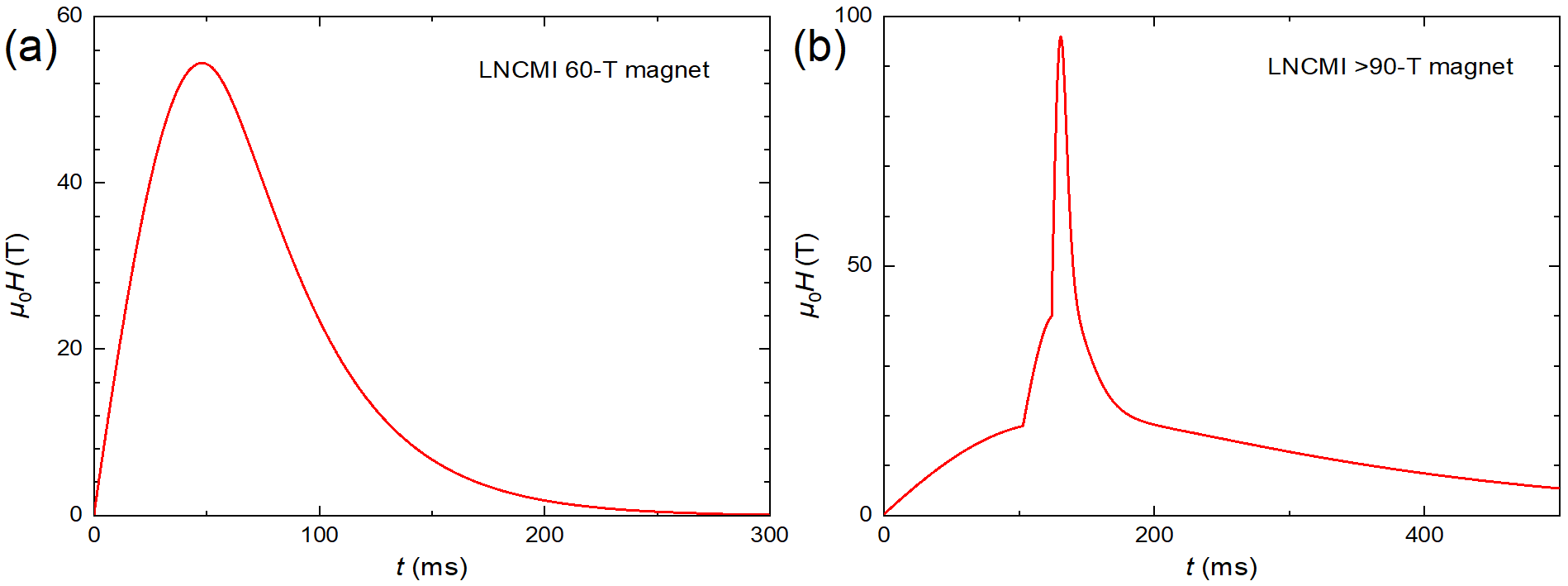}
\caption{\label{FigS1} Time profiles of magnetic fields generated by the (a) 60-T single-coil magnet and (b) $>90$-T triple-coil magnet used in this study.}
\end{figure*}

Four UTe$_2$ single crystals (samples $\#16$, $\#18$, $\#19$ and $\#20$) have been investigated here by electrical resistivity. In complement to the paper, which presents data from samples $\#18$ and $\#20$, this Supplemental Material shows data collected from the three samples $\#16$, $\#18$ and $\#19$, supporting the reproducibility of our results. The four samples have been grown by the molten-salt-flux (MSF) method \cite{Sakai2022s}. Their electrical resistivity was measured with a current $\mathbf{I}\parallel\mathbf{a}$ under magnetic fields up to 55~T applied in the (\textbf{b},\textbf{c}) plane for samples $\#16$, $\#18$ and $\#19$ and under magnetic fields $\mu_0\mathbf{H}\parallel\mathbf{b}$ up to 95~T for sample $\#20$. A single-axis rotator was used to change the field direction defined by the angle $\theta=(\mathbf{H},\mathbf{b})$ (with $\mathbf{H}\perp\mathbf{a}$). The samples $\#18$, $\#19$ and $\#20$ were selected with a clean face perpendicular to $\mathbf{c}$, with dimensions of approximately 1-2~mm along $\mathbf{a}$, 0.3-0.5~mm along $\mathbf{b}$, and 0.1-0.2~mm along $\mathbf{c}$. The sample $\#16$ was selected with a clean face perpendicular to the plane of Miller indices $(0 1 1)$, with similar dimensions than samples $\#18$ and $\#19$. The orientation of the crystals was ensured by Laue diffraction. Four electrical contacts with 15-$\mu$m gold wires were spot-welded on sample surfaces. The electrical resistivity was measured with an excitation current of 10~mA and a frequency $f\approx$~40~kHz. Electrical resistivity data with $\mathbf{I}\parallel\mathbf{a}$ were normalized so that the maximum value $\rho_{xx}=450~\rm{\mu\Omega cm}$ is reached at the temperature $T \simeq65$~K (the same normalization was used for our data published in \cite{Knafo2021as}). We note that a normalization coefficient leading to $\rho_{xx}=650~\rm{\mu\Omega cm}$ at $T \simeq65$~K was used in our first study published in \cite{Knafo2019s}, ending in overestimated values of the coefficient $A$ in this first work. Pulsed magnetic-field experiments were performed using a long-duration 60-T single-coil magnet (pulses rise of 50~ms and fall of 280~ms) and a $>90$-T triple-coil magnet (inner coil permitting to reach the maximum magnetic field: pulses rise of 6~ms and fall of 13~ms) at the Laboratoire National des Champs Magn\'{e}tiques Intenses (LNCMI) in Toulouse (see Figure \ref{FigS1}). Samples $\#16$, $\#18$ and $\#19$ were investigated at temperatures from 470~mK to 10~K delivered by an $^3$He insert combined with a $^4$He cryostat, and sample $\#20$ was investigated at temperatures from 1.4~K to 36.5~K using a $^4$He cryostat.

\subsection{Electrical resistivity and phase diagrams}

\begin{figure*}[t]
\centering
\includegraphics[width=0.85\textwidth]{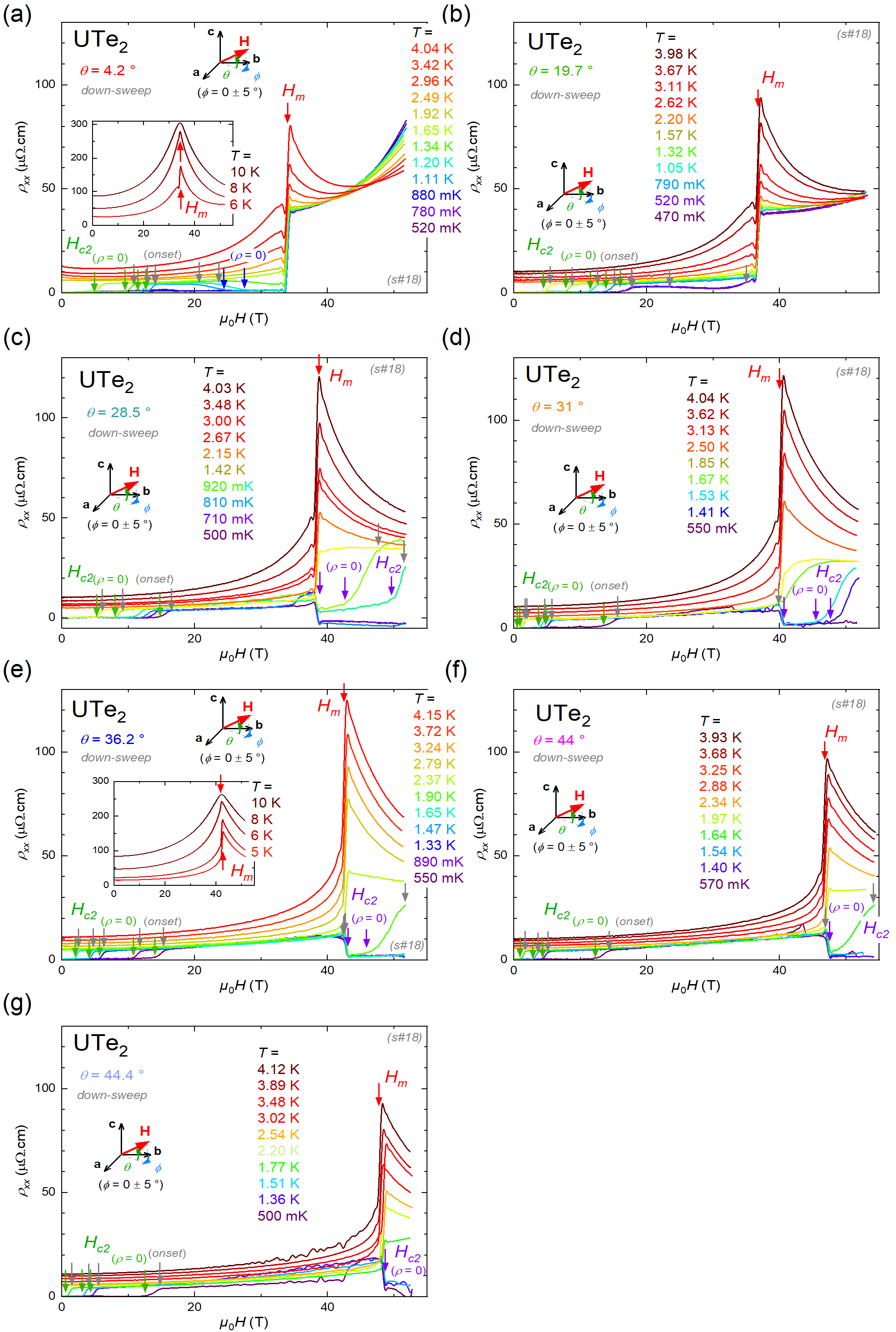}
\caption{\label{FigS2} Electrical resistivity $\rho_{xx}$ of UTe$_2$ sample $\#18$ versus magnetic field up to 55~T and indication of the characteristic transition fields, at temperatures $T$ from 500~mK to 10~K for field directions with the angles $\theta=$ (a) 4.2~$^\circ$, (b) 19.7~$^\circ$, (c) 28.5~$^\circ$, (d) 31~$^\circ$, (e) 36.2~$^\circ$, (f) 44~$^\circ$, and (g) 44.4~$^\circ$. Data from the down-sweep of the pulsed magnetic field are shown.}
\end{figure*}

\begin{figure*}[t]
\centering
\includegraphics[width=0.85\textwidth]{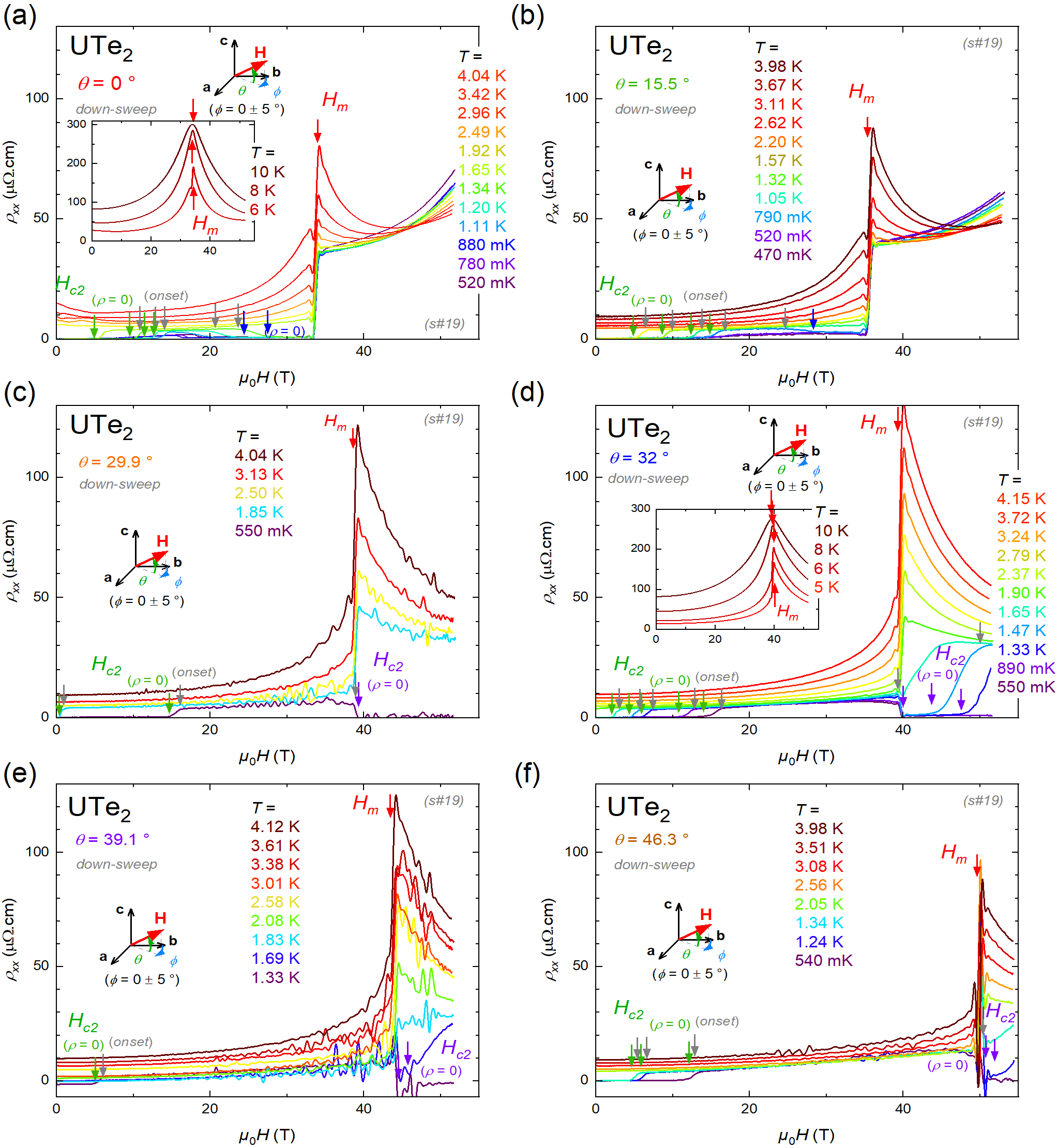}
\caption{\label{FigS3} Electrical resistivity $\rho_{xx}$ of UTe$_2$ sample $\#19$ versus magnetic field up to 55~T and indication of the characteristic transition fields, at temperatures $T$ from 500~mK to 10~K, for field directions with the angles $\theta=$ (a) 0~$^\circ$, (b) 15.5~$^\circ$, (c) 29.9~$^\circ$, (d) 32~$^\circ$, (e) 39.1~$^\circ$, and (f) 46.3~$^\circ$. Data from the down-sweep of the pulsed magnetic field are shown.}
\end{figure*}

\begin{figure*}[t]
\centering
\includegraphics[width=0.85\textwidth]{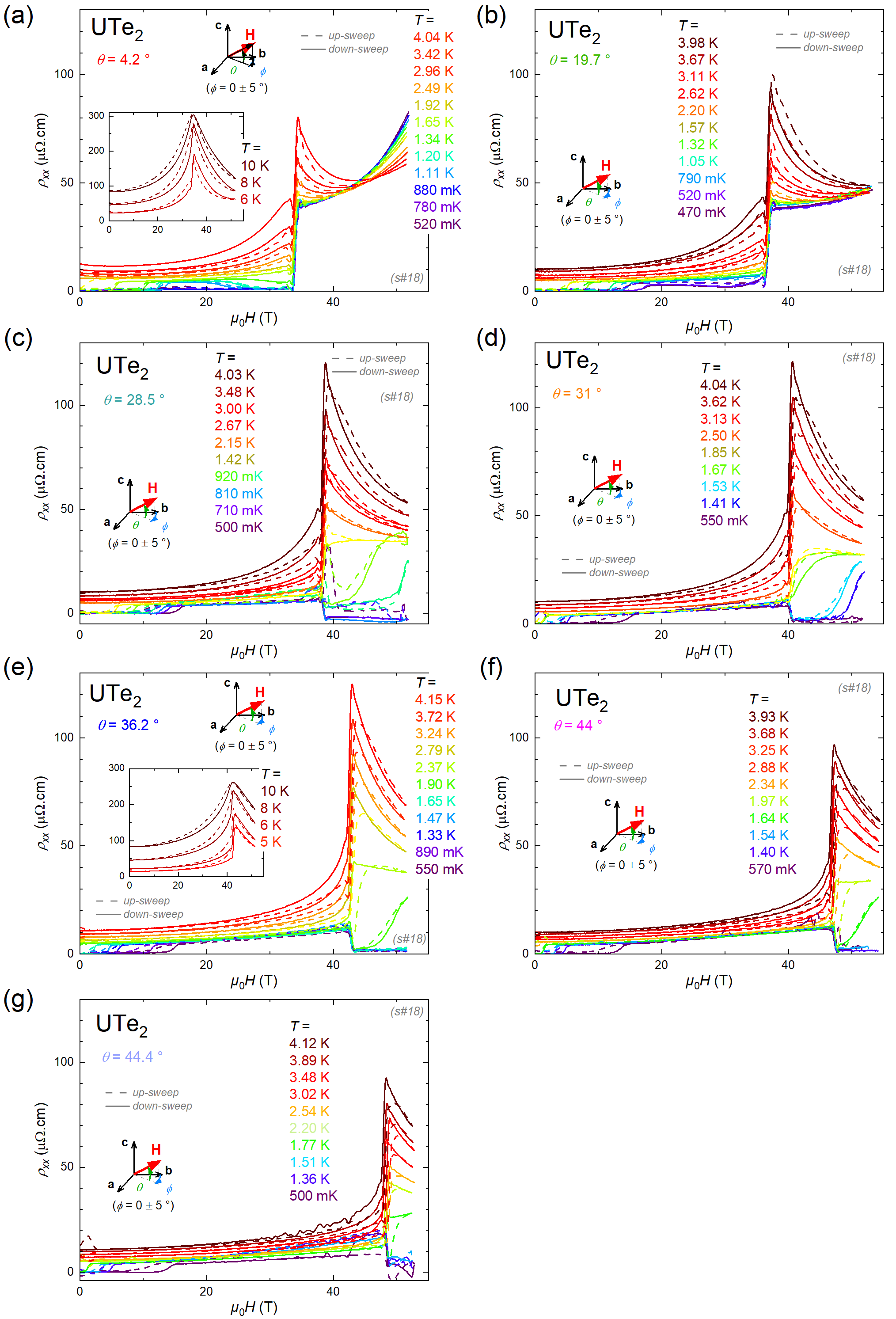}
\caption{\label{FigS4} Electrical resistivity $\rho_{xx}$ of UTe$_2$ sample $\#18$ versus magnetic field up to 55~T, at temperatures $T$ from 500~mK to 10~K, for field directions with the angles $\theta=$ (a) 4.2~$^\circ$, (b) 19.7~$^\circ$, (c) 28.5~$^\circ$, (d) 31~$^\circ$, (e) 36.2~$^\circ$, (f) 44~$^\circ$, and (g) 44.4~$^\circ$. Data from the up- and down-sweeps of the pulsed magnetic field are shown.}
\end{figure*}

\begin{figure*}[t]
\centering
\includegraphics[width=0.85\textwidth]{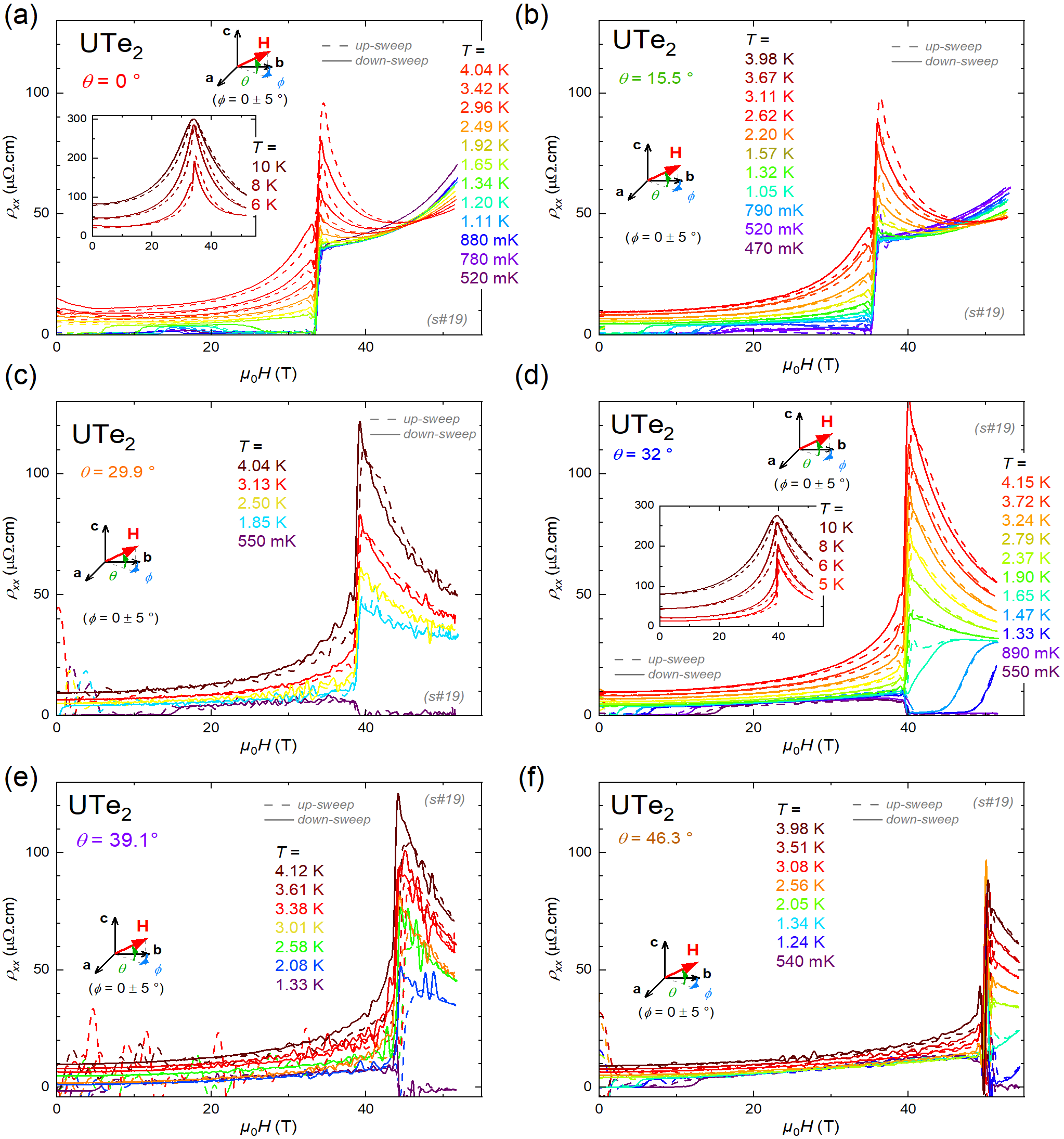}
\caption{\label{FigS5} Electrical resistivity $\rho_{xx}$ of UTe$_2$ sample $\#19$ versus magnetic field up to 55~T, at temperatures $T$ from 500~mK to 10~K, for field directions with the angles $\theta=$ (a) 0~$^\circ$, (b) 15.5~$^\circ$, (c) 29.9~$^\circ$, (d) 32~$^\circ$, (e) 39.1~$^\circ$, and (f) 46.3~$^\circ$. Data from the up- and down-sweeps of the pulsed magnetic field are shown.}
\end{figure*}

Figures \ref{FigS2} and \ref{FigS3} present the field variation of the electrical resistivity $\rho_{xx}$ of UTe$_2$ samples $\#18$ and $\#19$, respectively, at temperatures $T$ from 500~mK to 10~K, in magnetic fields up to 55~T with angles $\theta=(\mathbf{H},\mathbf{b})$ varying from $0~^\circ$ ($\mathbf{H}\parallel\mathbf{b}$) to $46~^\circ$. The critical magnetic fields at the metamagnetic and superconducting transitions are indicated for these data collected during the fall of the pulsed magnetic fields. Figures \ref{FigS4} and \ref{FigS5} present the field variation of the electrical resistivity $\rho_{xx}$ of UTe$_2$ samples $\#18$ and $\#19$, respectively, for both up and down-sweeps of the pulsed magnetic fields, for the same sets of temperatures and angles than in Figures \ref{FigS2} and \ref{FigS3}, respectively. They show the presence of hysteresis loops, which are stronger at low angles and low temperatures, and which result from the combination of magnetocaloric effect and self-heating of the samples by Eddy current, but also from the first-order nature of the metamagnetic transition. The two first effects have been minimized by reducing the dimensions (in particular the surface exposed to the magnetic field) and by an careful thermalization of the samples, but they could not be fully eliminated due to the fast variation of the magnetic field in our pulsed-field experiment. These loops are acceptable since they do not affect the analyzes and conclusions made here.

\begin{figure*}[htb]
\centering
\includegraphics[width=0.85\textwidth]{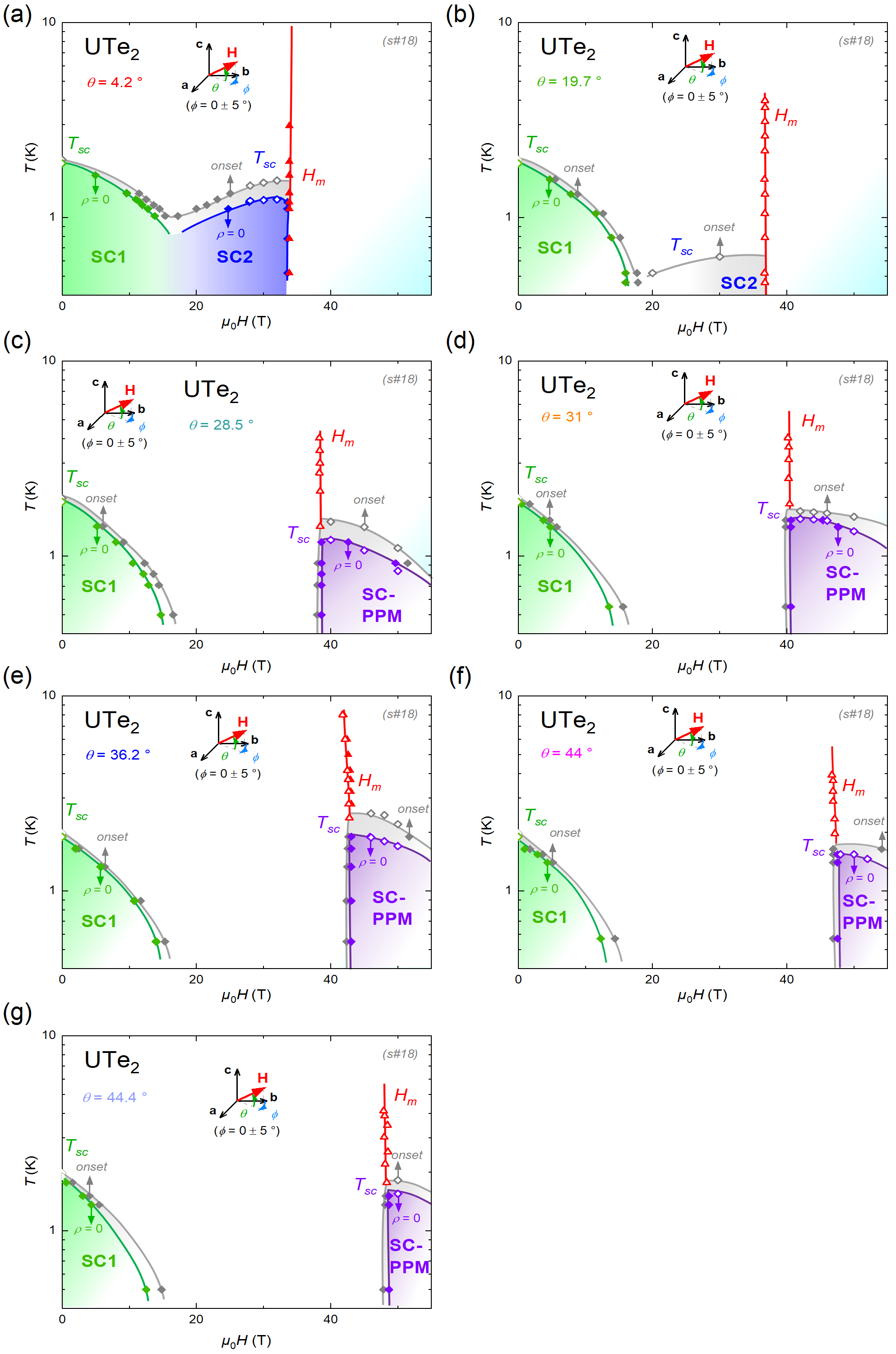}
\caption{\label{FigS6} Magnetic-field-temperature phase diagrams obtained from our resistivity measurements on sample $\#18$ for $\theta=$ (a) 4.2~$^\circ$, (b) 19.7~$^\circ$, (c) 28.5~$^\circ$, (d) 31~$^\circ$, (e) 36.2~$^\circ$, (f) 44~$^\circ$, and (g) 44.4~$^\circ$.}
\end{figure*}

\begin{figure*}[htb]
\centering
\includegraphics[width=0.85\textwidth]{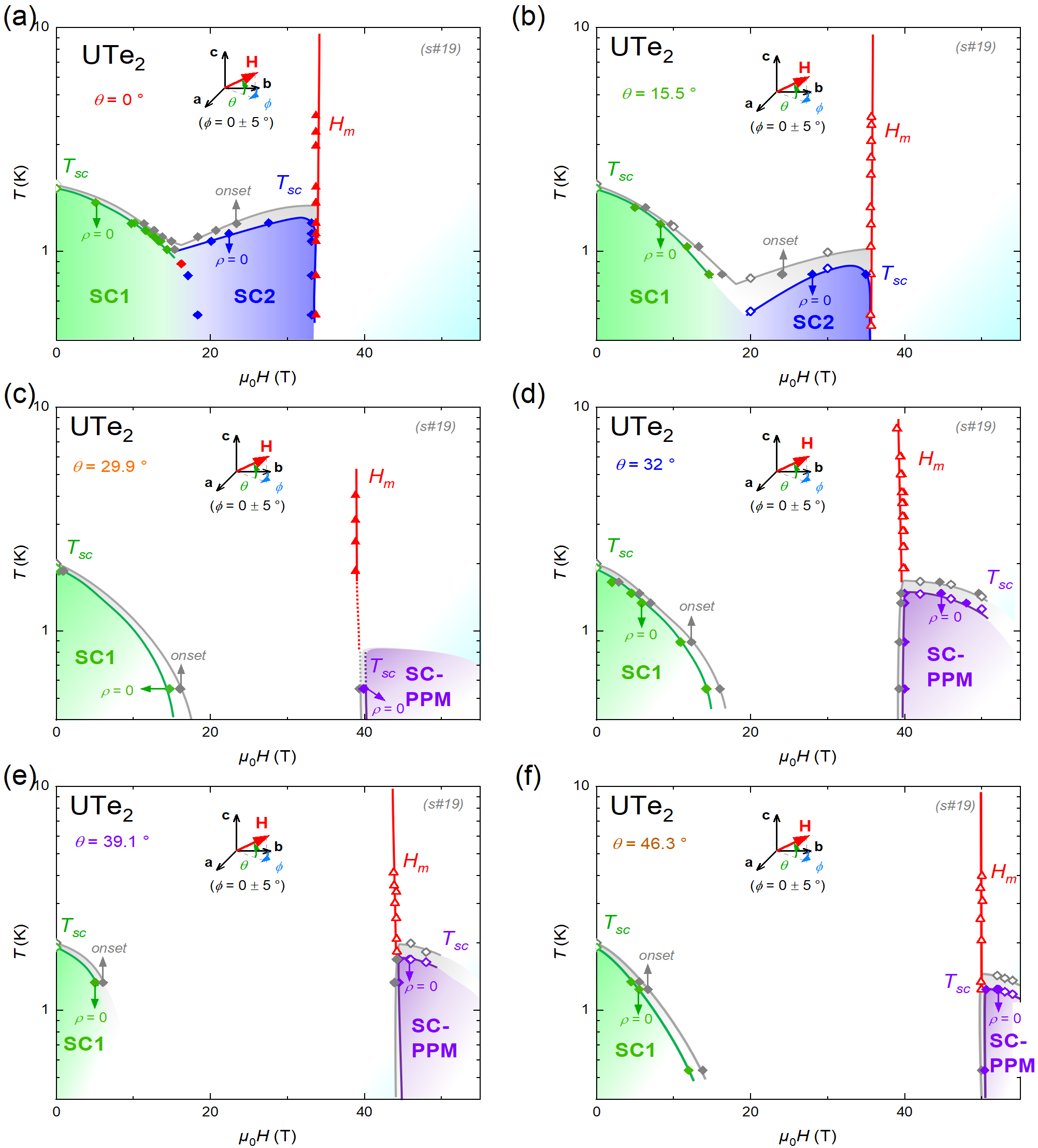}
\caption{\label{FigS7} Magnetic-field-temperature phase diagrams obtained from our resistivity measurements on sample $\#19$ for $\theta=$ (a) 0~$^\circ$, (b) 15.5~$^\circ$, (c) 29.9~$^\circ$, (d) 32~$^\circ$, (e) 39.1~$^\circ$, and (f) 46.3~$^\circ$.}
\end{figure*}

The magnetic-field temperature phase diagrams obtained from these resistivity data are presented in Figures \ref{FigS6} (sample $\#18$) and \ref{FigS7} (sample $\#19$).

Figure \ref{FigS8} shows the angle-magnetic-field phase diagram of UTe$_2$ obtained at low temperatures, combining data collected here on samples $\#18$ and $\#19$ and data published in the literature \cite{Knebel2019s,Ran2019bs,Knafo2021as,Schonemann2023s,Frank2024s,Wu2024s,Helm2024s,Lewin2025s}. Figure \ref{FigS9}(a,b) show intensity maps of the electrical resistivity of UTe$_2$ sample $\#18$ measured at $T=500$~mK and  $T=4$~K, respectively, as a function of $\theta$ and $H$. The phase diagram shown on top of the intensity maps was obtained from resistivity measurements on samples $\#18$ and $\#19$ at $T=500$~mK. Figure \ref{FigS9}(a) illustrates that the domains of stability of SC2 and SC-PPM correspond to zero-resistivity (or nearly-zero resistivity, see Section S4) states. Figure \ref{FigS9}(b) shows that an enhancement of $\rho_{xx}(T=4~\rm{K})$ is observed in the vicinity of SC-PPM, for $H\gtrsim H_m$ and angles $30\lesssim\theta\lesssim40~^\circ$. This enhancement of the electronic correlations visible at $T=4$~K is attributed to a magnetic-fluctuation mode, which also drives the enhancement of the Fermi-liquid coefficient $A$ in the same field and angle regime (see Figure 3 of the paper). This magnetic-fluctuation mode is suspected to be involved in the mechanism driving the superconducting phase SC-PPM.

\subsection{Fermi-liquid coefficient $A$}

Figures \ref{FigS10} and \ref{FigS11} present details about the low-temperature Fermi-liquid fits to the data by $\rho_{xx}=\rho_0+AT^2$ used to extract the coefficient $A$, for samples $\#18$ and $\#19$, respectively. The $T^2$ variation of $\rho_{xx}$ in magnetic fields up to 50~T and the fits to the data are presented for all directions of magnetic field investigated, with angles $\theta$ varying from $0~^\circ$ ($\mathbf{H}\parallel\mathbf{b}$) to $46~^\circ$.

For field directions close to $\mathbf{b}$, i.e., at low angles $\theta$, an increase of $\rho_{xx}$ at low temperature and high field is attributed to a carriers cyclotron-motion effect, which is induced in the highest-quality samples. This effect leads to a negative slope of $\rho_{xx}$ versus temperature at high magnetic fields. Figure \ref{FigS12} illustrates this by focusing on Fermi-liquid fits done on sample $\#18$ in a magnetic field applied with the angle $\theta=4.2~^\circ$. Figures \ref{FigS12} (a) and (b) present a comparison between Fermi-liquid fits to the data performed on different temperature windows, for up- and down-sweeps of the pulsed field, respectively. While fits done in the lowest-temperature window end in non-physical negative values of the coefficient $A$ at the largest fields, fits done at intermediate temperatures $T\geq2.4$~K, where the cyclotron-motion effect remains limited, permit to better estimate the variation of $A$ for $H>H_m$. Figure \ref{FigS12}(c) presents the field variation of $A$ extracted with the two sets of temperature windows on sample $\#18$ with the angle $\theta=4.2~^\circ$. For comparison, the field variation of $A$ extracted in a previous study on samples $\#1$, $\#2$ and $\#3$ in a magnetic field $\mathbf{H}\parallel\mathbf{b}$, which have been grown by the chemical vapor transport (CVT) method (data from \cite{Knafo2019s}) is also shown. In comparison with the MSF sample $\#18$, the CVT samples $\#1$, $\#2$ and $\#3$ have a lower quality, as indicated by a lower residual-resistivity ratio \cite{Aoki2024s} and a smaller cyclotron-motion effect. The CVT samples do not show a cyclotron-motion contribution to $\rho_{xx}$ at high fields and low temperatures, and they can be considered as a reference for a coefficient $A$ not affected by the cyclotron-motion effect. Figure \ref{FigS12}(c) shows that the variations of $A$ in magnetic fields $H>H_m$ extracted from intermediate-temperature fits on sample $\#18$ and extracted from low-temperature fits on samples $\#1$, $\#2$ and $\#3$ are very similar. This supports that the intermediate-temperature fits permit extracting a coefficient $A$, which is weakly affected by the cyclotron-motion effect on our MSF samples for low angles $\theta$ and $H>H_m$. For $H\lesssim H_m$, the coefficient $A$ extracted here for sample $\#18$ is smaller than that extracted on samples $\#1$, $\#2$ and $\#3$ in \cite{Knafo2019s}. This low-field difference is a consequence from a finer analysis allowed by smaller temperatures steps in the present study.

Figures \ref{FigS13} and \ref{FigS14} show the magnetic-field variations of $A$ extracted here for sample $\#18$ at angles 4.2~$^\circ\leq\theta\leq44.4^\circ$ and sample $\#19$ at angles 0~$^\circ\leq\theta\leq46.3^\circ$, respectively. Similar results are obtained for samples $\#18$ and $\#19$. At low angles $\theta<20^\circ$, the alternative low-temperature fits, for which a deviation - due to the cyclotron-motion effect - of $A$ towards negative values is induced beyond $H_m$ are also shown in the panels (a) and (b). For angles $\theta >20~^\circ$, no cyclotron motion is detected and the two types of fits give similar results. These graphs show small hysteresis loops in the variation of $A$, which are the consequence of the small loops observed in the variation of $\rho_{xx}$.

Figure \ref{FigS15} shows the magnetic-field variation of the residual resistivity $\rho_0$ extracted for samples $\#18$ at angles 4.2~$^\circ\leq\theta\leq44.4^\circ$ and sample $\#19$ at angles 0~$^\circ\leq\theta\leq46.3^\circ$. Similar results are obtained for samples $\#18$ and $\#19$. A jump in the residual resistivity occurs at the metamagnetic transition as previously reported in \cite{Knafo2019s,Knafo2021as,Knebel2024s}. For $\theta<20^\circ$, $\rho_0$ increases in high fields due to a contribution from the cyclotron-motion effect, whose signatures are not observed for $\theta>20~^\circ$.

Figure \ref{FigS16} shows the variation of $A$ extracted from Fermi-liquid fits to the electrical resistivity of UTe$_2$ sample $\#18$, for 4.2~$^\circ\leq\theta\leq44.4~^\circ$ (from high-$T$ fits for $\theta<20~^\circ$ and low-$T$ fits for $\theta>20~^\circ$) as a function of $H/H_m$. It emphasizes that the variation of $A$ with magnetic field becomes more asymmetric through $H_m$ at angles $\theta<20~^\circ$, with a shoulder indicative of enhanced correlations for $H>H_m$.

\begin{figure*}[hb]
\centering
\includegraphics[width=0.42\textwidth]{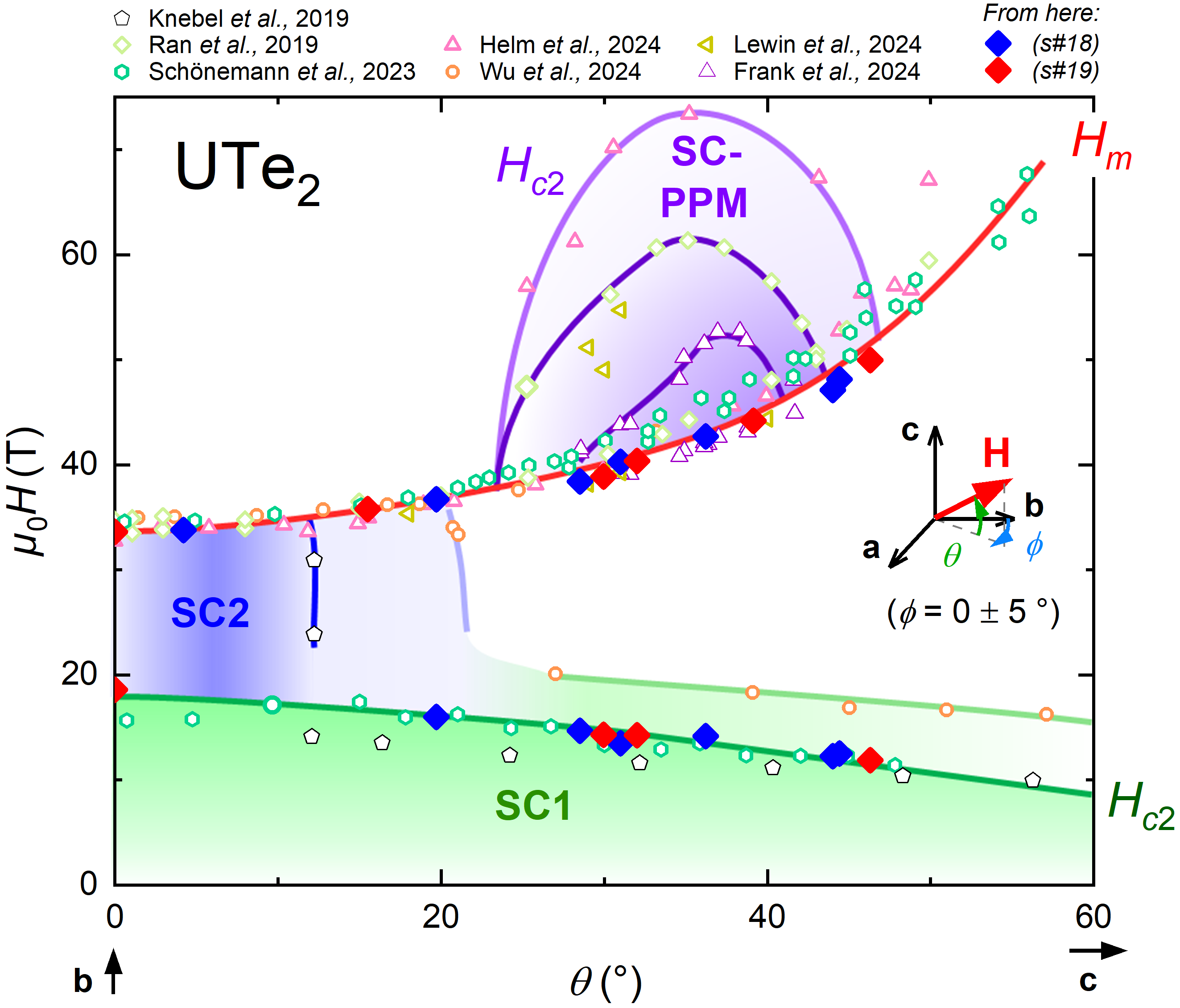}
\caption{\label{FigS8} Angle-magnetic-field phase diagram of UTe$_2$ in the limit of $T\rightarrow0$, from a compilation of data collected here on samples $\#18$ and $\#19$ and from \cite{Knebel2019,Ran2019b,Knafo2021a,Schonemann2023,Frank2024,Wu2024,Helm2024,Lewin2025}}
\end{figure*}

\begin{figure*}[tb]
\centering
\includegraphics[width=.95\textwidth]{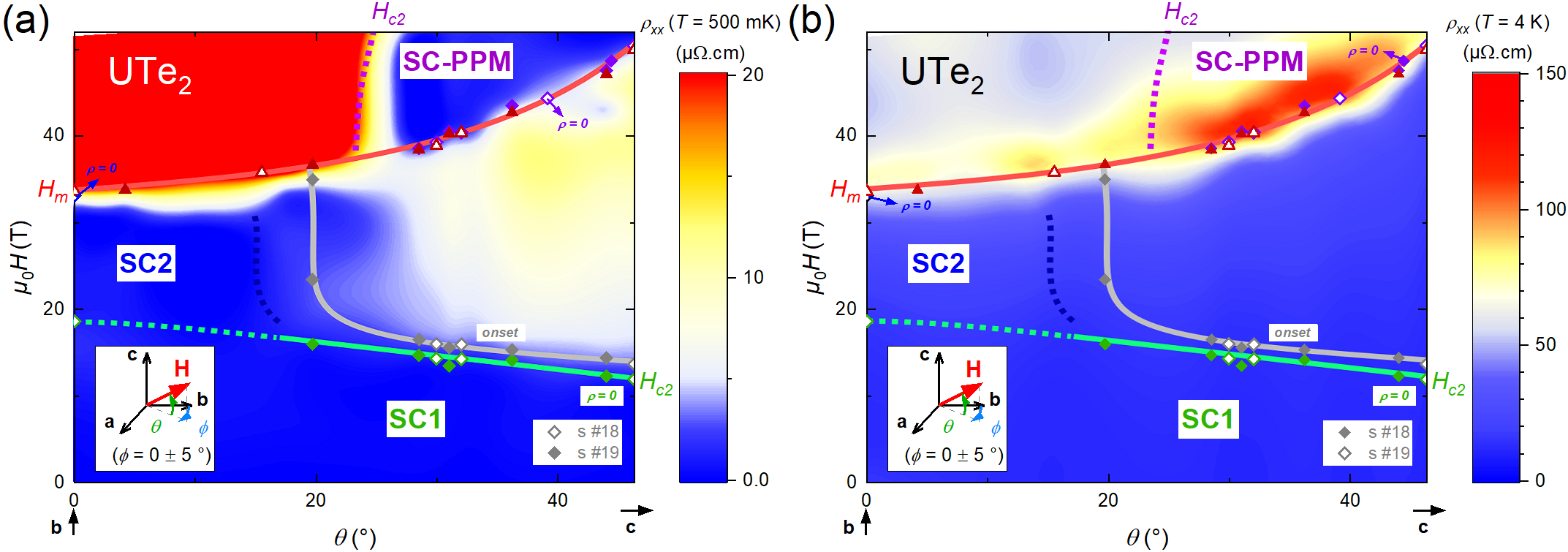}
\caption{\label{FigS9} Intensity maps of the electrical resistivity of UTe$_2$ sample $\#18$ measured at (a) $T=500$~mK and (b) $T=4$~K, as a function of $\theta=(\mathbf{b},\mathbf{H})$ and $H$ (for $\mathbf{H}\perp\mathbf{a}$). The phase diagram shown on top of the intensity maps was obtained from resistivity measurements on samples $\#18$ and $\#19$ at $T=500$~mK.The dashed lines indicate expected transition lines which were not probed here.}
\end{figure*}

\begin{figure*}[htb]
\centering
\includegraphics[width=0.85\textwidth]{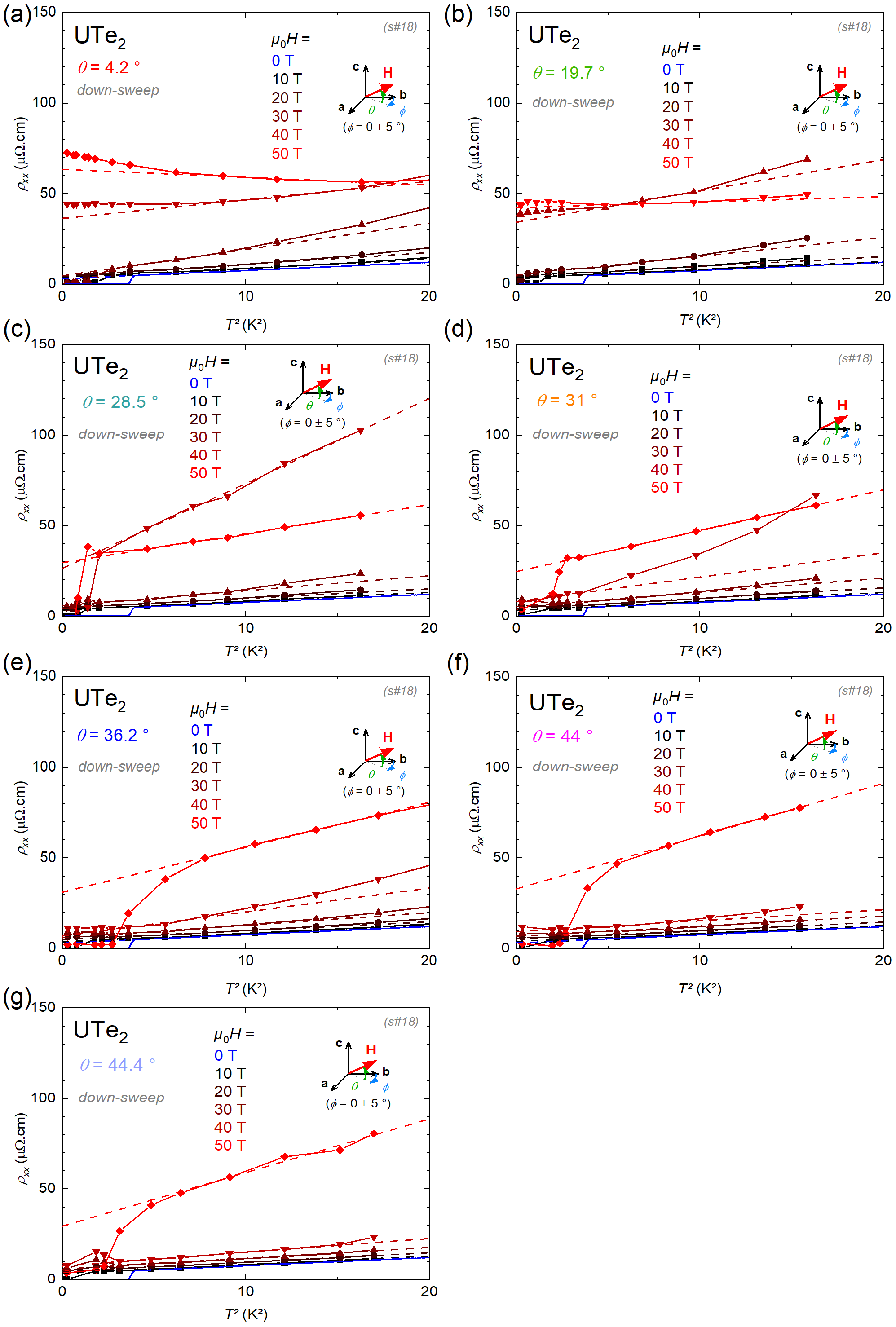}
\caption{\label{FigS10} Electrical resistivity $\rho_{xx}$ versus $T^2$ of UTe$_2$ sample $\#18$ at magnetic fields $\mu_0H=0$, 10, 20, 30, 40, and 50~T, for $\theta=$ (a) 4.2~$^\circ$, (b) 19.7~$^\circ$, (c) 28.5~$^\circ$, (d) 31~$^\circ$, (e) 36.2~$^\circ$, (f) 44~$^\circ$, and (g) 44.4~$^\circ$. The dashed lines correspond to Fermi-liquid fits by $\rho_{xx}=\rho_0+AT^2$.}
\end{figure*}

\begin{figure*}[htb]
\centering
\includegraphics[width=0.85\textwidth]{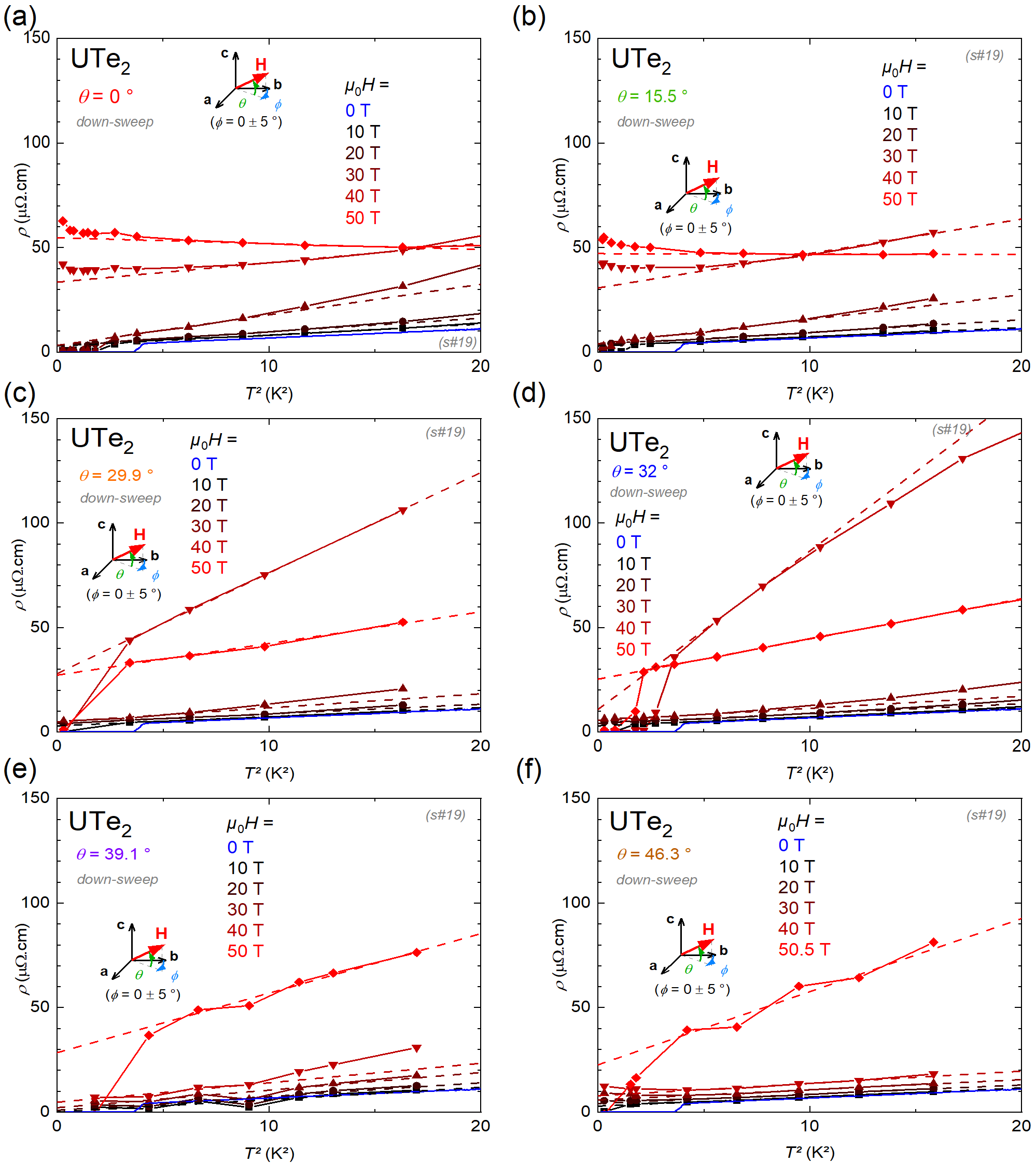}
\caption{\label{FigS11} Electrical resistivity $\rho_{xx}$ versus $T^2$ of UTe$_2$ sample $\#19$ at magnetic fields $\mu_0H=0$, 10, 20, 30, 40, and 50~T, for $\theta=$ (a) 0~$^\circ$, (b) 15.5~$^\circ$, (c) 29.9~$^\circ$, (d) 32~$^\circ$, (e) 39.1~$^\circ$, and (f) 46.3~$^\circ$. The dashed lines correspond to Fermi-liquid fits by $\rho_{xx}=\rho_0+AT^2$.}
\end{figure*}

\begin{figure*}[tb]
\centering
\includegraphics[width=0.8\textwidth]{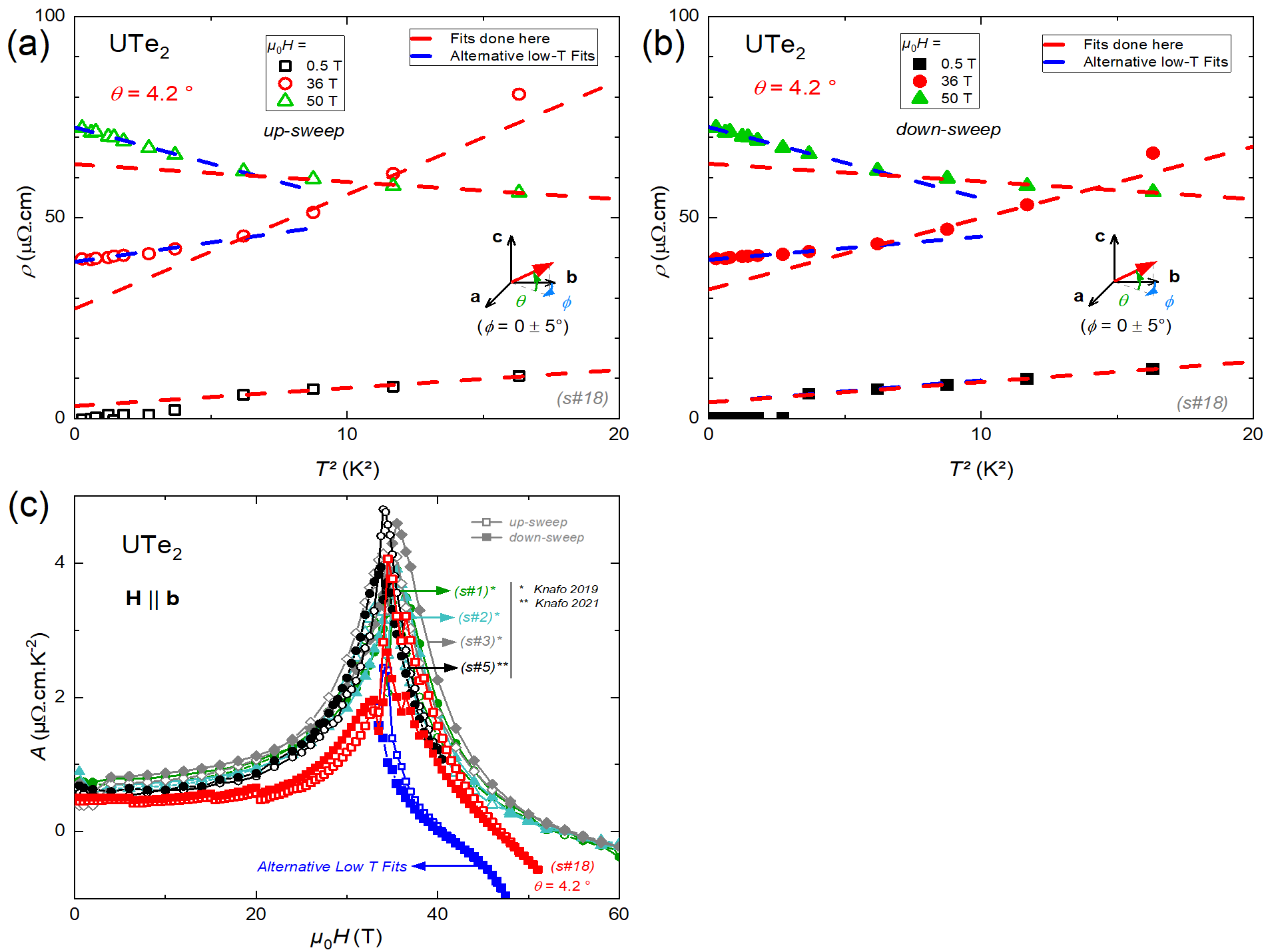}
\caption{\label{FigS12} Electrical resistivity $\rho_{xx}$ versus $T^2$ of UTe$_2$ sample $\#18$ at magnetic fields $\mu_0H=0.5$, 36, and 50~T, for $\theta=4.2^\circ$ shown for (a) the up-sweep and (b) the down-sweep of the magnetic-field pulses. The red dashed lines correspond to Fermi-liquid fits by $\rho_{xx}=\rho_0+AT^2$ done in this work, and the blue dashed lines show alternative low-temperature fits to high-field data, for which a low-temperature cyclotron-motion-driven enhancement of $\rho_{xx}$ at low temperature ends in non-physical negative values of $A$. (c) Magnetic-field dependence of the Fermi-liquid coefficient $A$ extracted on sample $\#18$ grown by the MSF method with $\theta=4.2^\circ$ (fit done here in red and alternative low-$T$ fits in blue) and on samples $\#1$, $\#2$ and $\#3$ grown by the CVT method and studied with $\mathbf{H}\parallel\mathbf{b}$ in \cite{Knafo2019} (low-$T$ fits).}
\end{figure*}

\begin{figure*}[tb]
\centering
\includegraphics[width=0.85\textwidth]{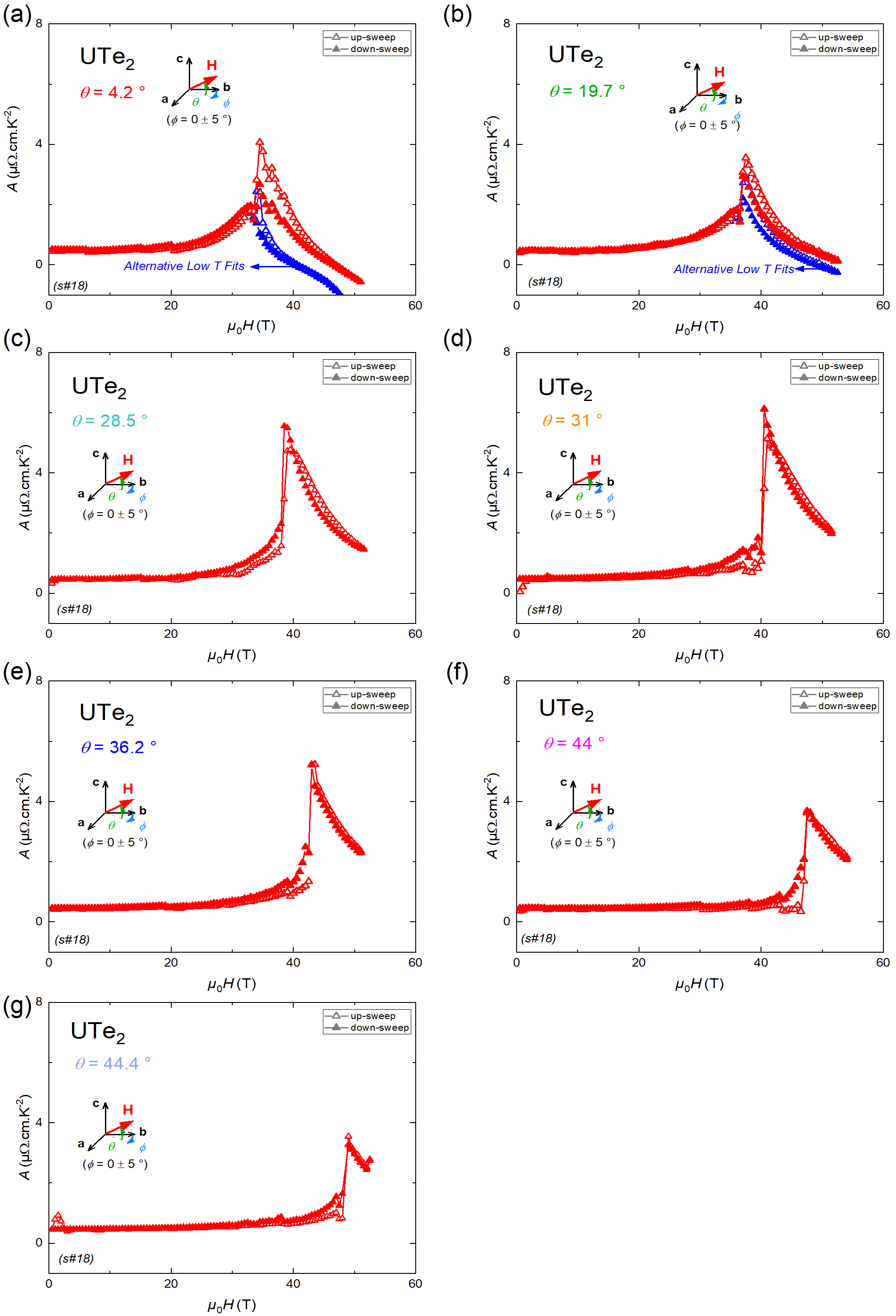}
\caption{\label{FigS13} Magnetic-field variation of the Fermi-liquid coefficient $A$ extracted here (red) and from alternative low-$T$ fits, done at high fields and low angles $\theta$, (blue) to the electrical resistivity $\rho_{xx}$ of UTe$_2$ sample $\#18$, for $\theta=$ (a) 4.2~$^\circ$, (b) 19.7~$^\circ$, (c) 28.5~$^\circ$, (d) 31~$^\circ$, (e) 36.2~$^\circ$, (f) 44~$^\circ$, and (g) 44.4~$^\circ$. Data from the up- and down-sweeps of the pulsed magnetic field are shown.}
\end{figure*}

\begin{figure*}[tb]
\centering
\includegraphics[width=0.85\textwidth]{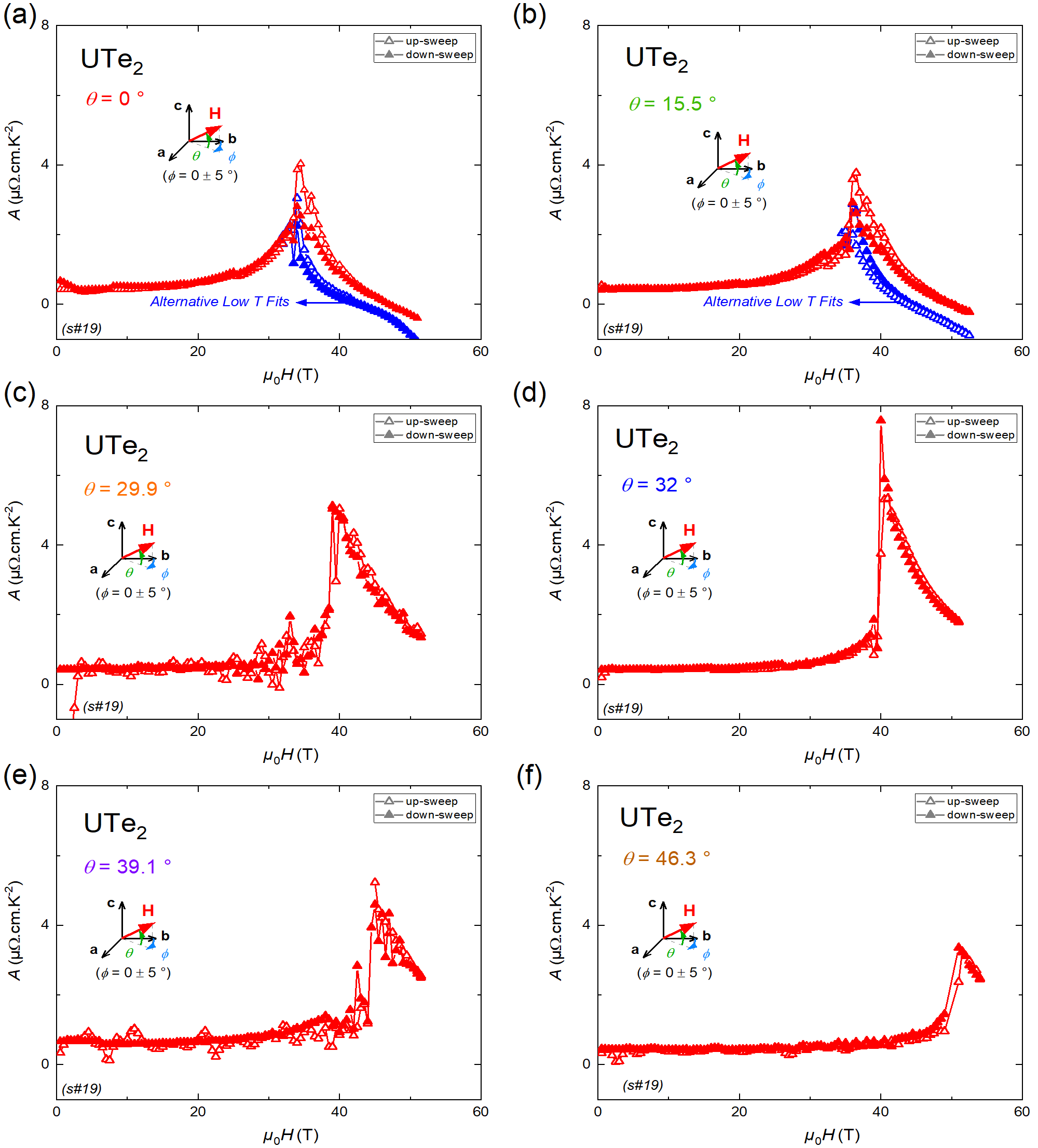}
\caption{\label{FigS14} Magnetic-field variation of the Fermi-liquid coefficient $A$ extracted here (red) and from alternative low-$T$ fits, done at high fields and low angles $\theta$, (blue) to the electrical resistivity $\rho_{xx}$ of UTe$_2$ sample $\#19$, for $\theta=$ (a) 0~$^\circ$, (b) 15.5~$^\circ$, (c) 29.9~$^\circ$, (d) 32~$^\circ$, (e) 39.1~$^\circ$, and (f) 46.3~$^\circ$. Data from the up- and down-sweeps of the pulsed magnetic field are shown.}
\end{figure*}

Figure \ref{FigS17} compares the magnetic-field variations of $A$ extracted for different angles $\theta$ on the three samples ($\#16$, $\#18$, and $\#19$) grown by the MSF technique investigated here and on other samples ($\#1$, $\#2$, $\#3$, $\#5$, $\#6$, and $\#7$) grown by the CVT technique and investigated in \cite{Knafo2019s,Knafo2021as}. Samples $\#18$ and $\#19$ have been studied in angles $0\leq\theta\leq46~^\circ$ and sample $\#16$ has been studied in angles $30\leq\theta\leq46~^\circ$. Samples $\#1$, $\#2$, $\#3$ and $\#5$  have been studied in a magnetic field $\mathbf{H}\parallel\mathbf{b}$ with $\theta=0~^\circ$. Samples $\#6$, and $\#7$ have also been studied in a magnetic field tilted by an angle initially estimated by $\theta=27\pm5~^\circ$ using a sample holder with a fixed tilted direction. However, different values of the metamagnetic field indicate different angles, which were re-estimated here as $\theta=40~^\circ$ for sample $\#6$, and $\theta=40.7$ and $41~^\circ$ for sample $\#7$, assuming no tilt of the magnetic field towards the direction $\mathbf{a}$, from the now-well-characterized angle-$\theta$ dependence of $H_m$  (see Figure \ref{FigS8} and \cite{Ran2019bs,Schonemann2023s,Frank2024s,Wu2024s,Helm2024s,Lewin2025s}). The data compiled in Figure \ref{FigS17} show similar variations of $A$ as a function of $H$ and $\theta$ and demonstrate the reproducibility of the results presented in our paper. The comparison of the data extracted for the different samples indicate a typical error of about $\pm0.5~\rm{\mu\Omega cm K}^{-2}$ in the determination of $A$. The maximum of $A$ at $H_m$ is enhanced at angles theta between 30 and $35~^\circ$, but it has similar values for $\theta=0$ and $40~^\circ$, in agreement with the data published in \cite{Knafo2021as}. In the present study, we have also increased the number of investigated temperatures, which permitted extracting a more pronounced asymmetry of the maximum of $A$ at $H_m$ for large angles than in our previous work \cite{Knafo2021as}. A small hysteresis is visible in all sets of data and results from different effects: heating by Eddy currents, magnetocaloric effect, and the presence of a first-order metamagnetic transition. The hysteresis is found here to be a bit larger at low angles $\theta$ for the MSF samples, probably in relation with the anisotropic electronic properties of UTe$_2$. A smaller hysteresis is observed in the previously studied CVT samples, which may be a consequence of the larger low-temperature electrical resistivity of the CVT samples.

\subsection{Comments on Fermi-liquid fits}

\begin{figure*}[tb]
\centering
\includegraphics[width=0.85\textwidth]{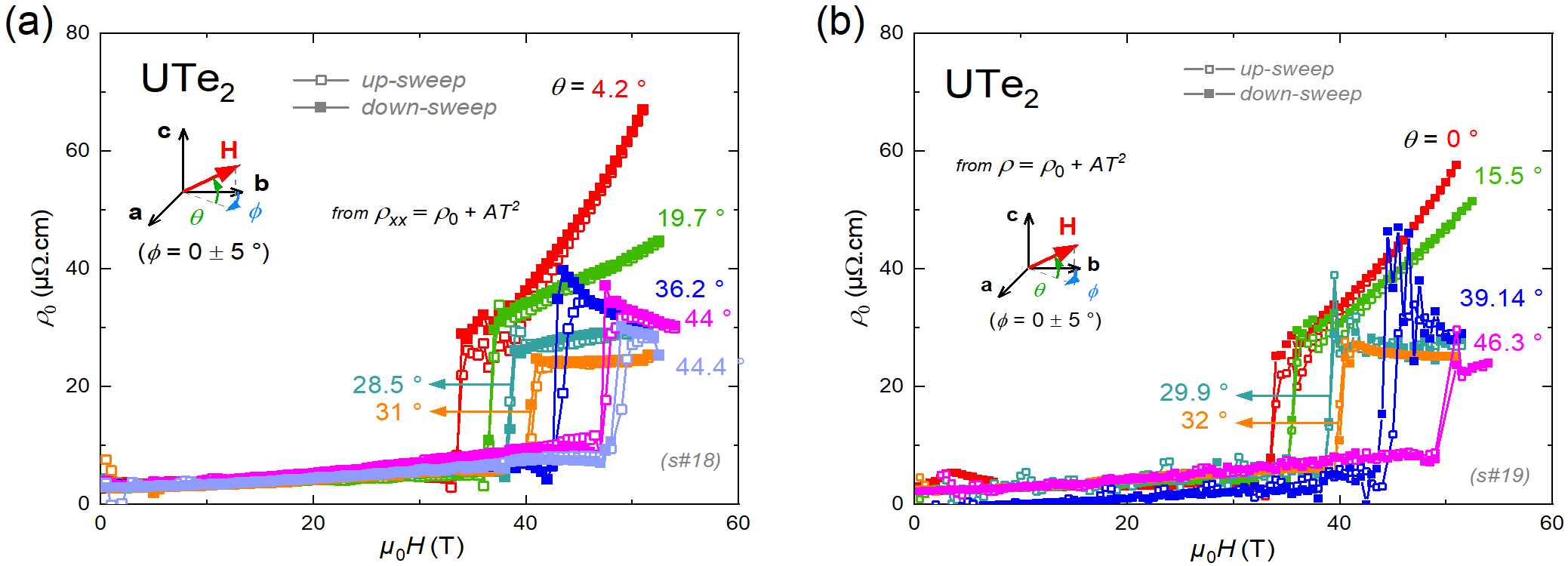}
\caption{\label{FigS15} Magnetic-field variation of the residual resistivity $\rho_0$ extracted from Fermi-liquid fits to the electrical resistivity of UTe$_2$ (a) sample $\#18$ for 4.2~$^\circ\leq\theta\leq44.4^\circ$ and (b) sample $\#19$ for 0~$^\circ\leq\theta\leq46.3^\circ$. Data from the up- and down-sweeps of the pulsed magnetic field are shown.}
\end{figure*}

\begin{figure*}[tb]
\centering
\includegraphics[width=0.42\textwidth]{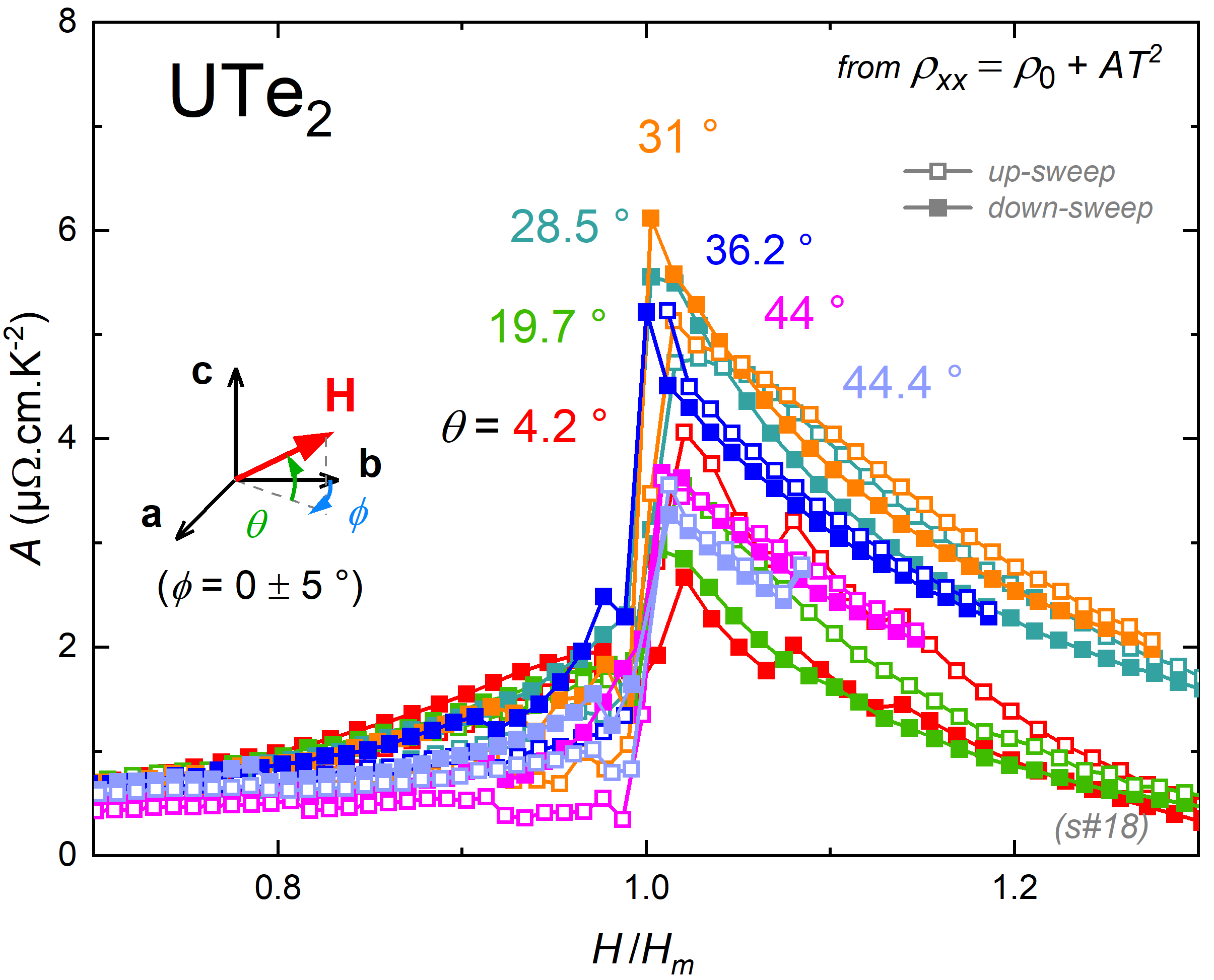}
\caption{\label{FigS16} Variation of the coefficient $A$ extracted from Fermi-liquid fits to the electrical resistivity of UTe$_2$ sample $\#18$, for 4.2~$^\circ\leq\theta\leq44.4~^\circ$, as a function of $H/H_m$. Data from the up- and down-sweeps of the pulsed magnetic field are shown.}
\end{figure*}

\begin{figure*}[tb]
\centering
\includegraphics[width=0.8\textwidth]{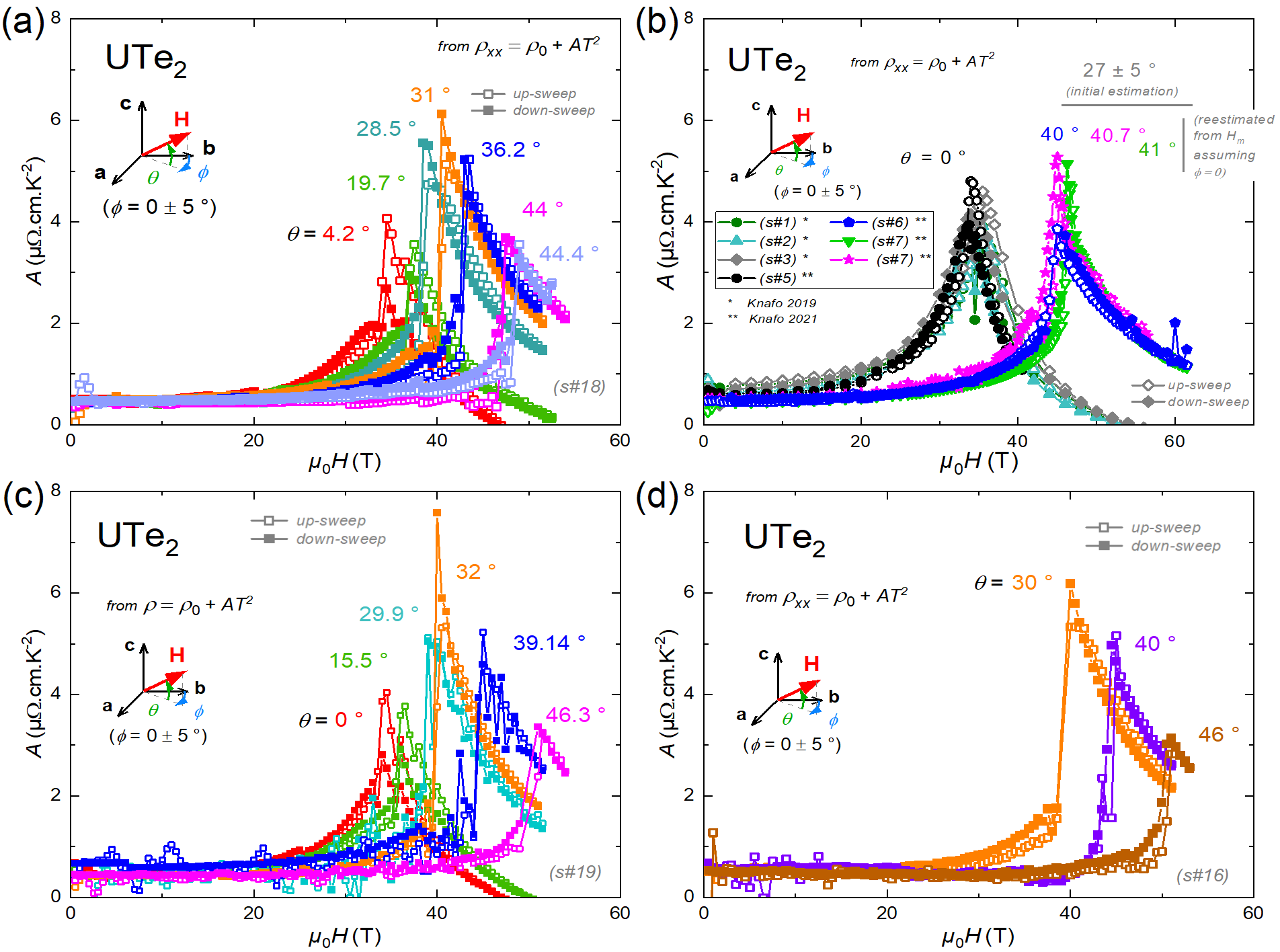}
\caption{\label{FigS17} Magnetic-field variation of $A$ obtained from Fermi-liquid fits to the electrical resistivity $\rho_{xx}$ of (a) sample $\#18$ for 4.2~$^\circ\leq\theta\leq44.4^\circ$, (b) samples $\#1$ ($\theta=0^\circ$), $\#2$ ($\theta=0^\circ$), $\#3$ ($\theta=0^\circ$), $\#5$ ($\theta=0^\circ$), $\#6$ ($\theta=40^\circ$), and $\#7$ ($\theta=40.7^\circ$ and $41^\circ$), (c) sample $\#19$ for 0~$^\circ\leq\theta\leq46.3^\circ$, and (d) $\#16$ for 30~$^\circ\leq\theta\leq46^\circ$. Samples $\#16$, $\#18$, and $\#19$ have been grown by the MSF technique and studied here, and samples $\#1$, $\#2$, $\#3$, $\#5$, $\#6$, and $\#7$ have been grown by the CVT technique and investigated in \cite{Knafo2019,Knafo2021a}. Data from the up- and down-sweeps of the pulsed magnetic field are shown.}
\end{figure*}

Fermi-liquid fits are usually done in a low temperature window. In this work, we were sometimes obliged to make the fits in a narrow temperature window. The low-temperature points had not to be considered in the case of a superconducting phase or a cyclotron-motion effect, as well as higher-temperature data points has not to be considered when the $H_m(T)$ transition line was crossed [in fields $H\lesssim H_m(T\rightarrow0)$]. We have also checked the field-dependence of the extracted residual resistivity (shown in Figure \ref{FigS15}), for which non-physical negative values can indicate a non-appropriate temperature window for the fit (which needs then to be corrected). We recall that constructing $T$-dependent and constant-field resistivity curves from pulsed-field resistivity data collected at constant temperatures leads to more noisy data than those who would be directly measured as function of temperature in a steady field. A challenge in pulsed-field experiments is to reduce this source of noise in the data, and a particular care has to be given to the experimental conditions (the quality of electrical contacts is important to reduce the out-of-phase signal in high-frequency measurements, the temperature gradients in the cryostat have to be considered, possible heating effects have to be identified and reduced etc.).

At low-temperature, a Fermi-liquid behavior associated with an enhanced effective mass near a metamagnetic transition, as observed here or in \cite{Knafo2019s,Knafo2021as,Thebault2022s} in UTe$_2$ , is not a surprising result. It has been also observed in the electrical resistivity of other heavy-fermion paramagnets, as CeRu$_2$Si$_2$ \cite{Aoki2011bs} or CeRh$_2$Si$_2$ under pressure \cite{Knafo2017s}. An enhanced Fermi-liquid effective mass indicates the presence of enhanced low-energy magnetic fluctuations, generally associated with a lower temperature scale. It is related with quantum criticality at a quantum phase transition, which can be induced by a magnetic field, pressure, or chemical doping. A maximum of the Fermi-liquid coefficient and a maximum of the intensity of the magnetic fluctuations are generally observed. They are associated with a minimal and finite characteristic temperature (rather than a diverging intensity and a zero-temperature scale expected theoretically at a second-order quantum phase transition). Then, as long as the data were collected at temperatures below the characteristic temperature scale of the magnetic fluctuations, a Fermi-liquid description can be appropriate. Enhanced but not diverging magnetic fluctuations have for instance been observed microscopically at the quantum magnetic phase transition of Ce$_{1-x}$La$_x$Ru$_2$Si$_2$ \cite{Knafo2009s}.

\subsection{Flux-flow resistivity}

\begin{figure*}[tb]
\centering
\includegraphics[width=0.85\textwidth]{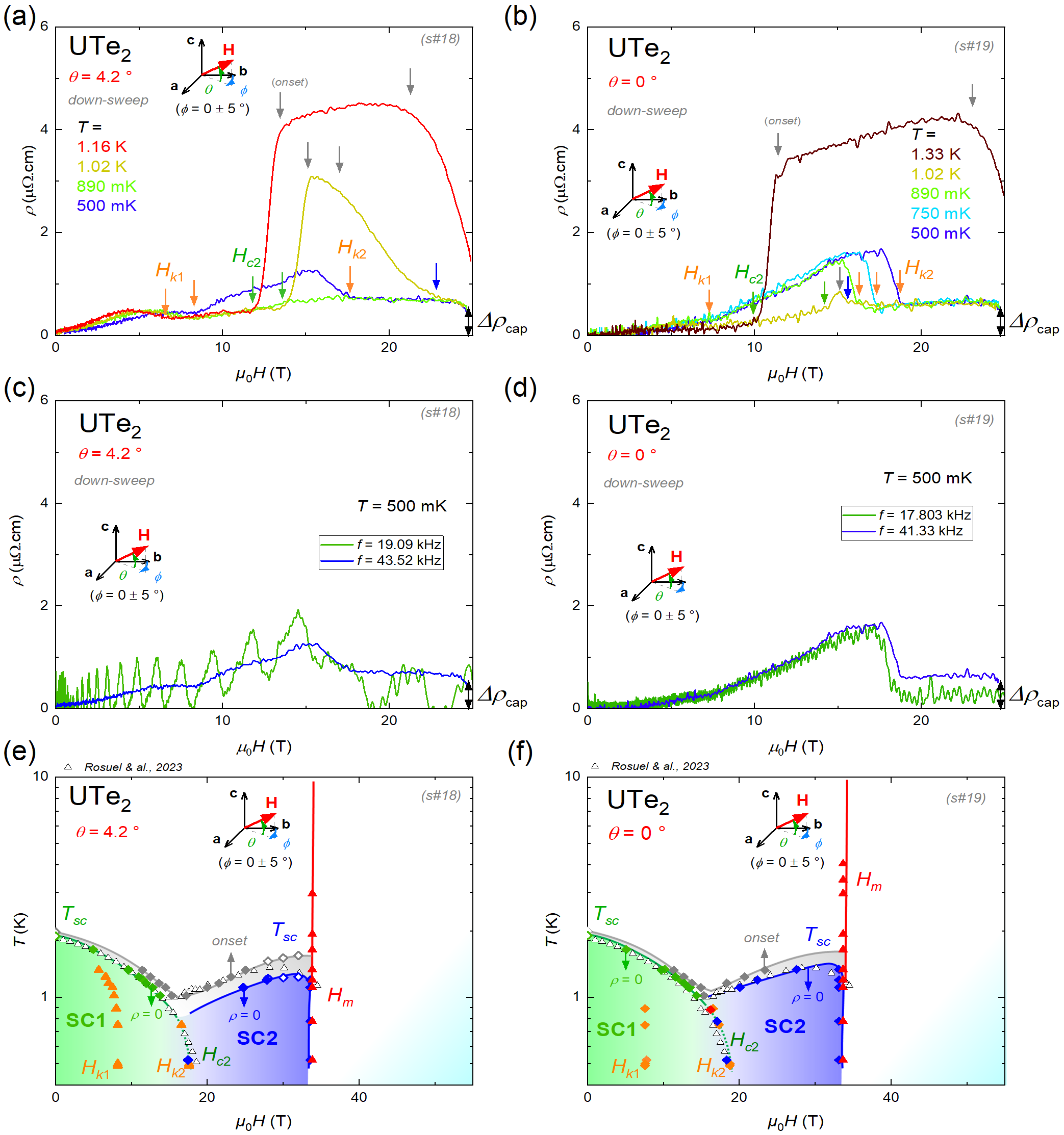}
\caption{\label{FigS18} Electrical resistivity versus magnetic field up to 25~T of (a) sample $\#18$ for $\theta=4.2^\circ$ at temperatures 500~mK~$\leq T\leq1.16$~K and (b) sample $\#19$ for $\theta=0^\circ$ at temperatures 500~mK~$\leq T\leq1.33$~K. Comparison of two pulsed-field shots performed at $T=500$~mK on (c) sample $\#18$ for $\theta=4.2^\circ$, at the current frequencies $f=19.09$ and 43.52~kHz and (d) sample $\#19$ for $\theta=0^\circ$, at the current frequencies $f=17.803$ and 41.33~kHz, showing the presence of a non-intrinsic contribution $\Delta\rho_{cap}\simeq0.5~\mu\Omega$ to the resistivity at 25~T induced by capacitive effects at frequencies $f\simeq40$~kHz. Magnetic-field-temperature phase diagrams determined from electrical-resistivity data collected here on (e) sample $\#18$ for $\theta=4.2^\circ$ and (f) sample $\#19$ for $\theta=0^\circ$. These phase diagrams include superconducting phase boundaries determined from heat-capacity measurements in \cite{Rosuel2023a}.}
\end{figure*}

Figures \ref{FigS18}(a-b) present a zoom on $\rho_{xx}$ measured on samples $\#18$ and $\#19$, respectively, at temperatures $T<1.4$~K and magnetic fields $\mu_0\mathbf{H}\approx\parallel\mathbf{b}$ up to 25~T. At the lowest temperatures, we find that, instead of a perfectly zero resistivity, a small contribution to the electrical resistivity is measured in the superconducting state. It results from the combination of two effects. The first one is a capacitive effect due to non-perfect electrical contacts on the sample, inducing a non-intrinsic contribution $\Delta\rho_{cap}\simeq0.5~\rm{\mu\Omega cm}$ to the electrical resistivity, measured here at a high frequency of $\simeq40$~kHz. As shown in Figures \ref{FigS18}(c-d), this capacitive effect is smaller at smaller frequencies $\simeq15-19$~kHz. The second effect results from a flux-flow contribution to the high-field electrical resistivity, which is observed here in the superconducting phase SC1, but not in the superconducting phase SC2. Two kinks observed in the electrical resistivity at $T=500$~mK at the magnetic fields $\mu_0H_{k1}$ and $\mu_0H_{k2}$, which reach 8.2~T and 17.6~T in sample $\#18$, and 7.5~T and 18.7~T in sample $\#19$, respectively, are attributed to the flux-flow contribution to $\rho_{xx}$. Figures \ref{FigS18}(e-f) present the magnetic-field-temperature phase diagrams obtained on samples $\#18$ and $\#19$, respectively, in magnetic fields $\mu_0\mathbf{H}\approx\parallel\mathbf{b}$. The temperature dependence of $H_{k2}$ coincides with that of the critical field $H_{c2}$ delimiting the superconducting phase SC1, which was determined by heat-capacity measurements in \cite{Rosuel2023as}. This indicates that SC1 is associated with a larger flux-flow resistivity than SC2. Flux-flow contributions to the electrical resistivity $\rho_{xx}$ of UTe$_2$ under a magnetic field $\mathbf{H}\approx\parallel\mathbf{b}$ have also been observed in \cite{Sakai2023,Tokiwa2023}.


%

\newpage

\end{document}